\documentstyle[12pt]{article}
\def\ssl#1{\rlap{\hbox{$\mskip 3 mu /$}}#1}  
\begin{document}


\def\Nequalstwo{\Psi}
\def\eff{{\rm eff}}
\def\inst{{\rm inst}}
\def\fermi{{\rm fermi}}
\def\trtwo{\tr^{}_2\,}
\def\finv{f^{-1}}
\def\Ubar{\bar U}
\def\wbar{\bar w}
\def\fbar{\bar f}
\def\abar{\bar a}
\def\bbar{\bar b}
\def\Deltabar{\bar\Delta}
\def\dalpha{{\dot\alpha}}
\def\dbeta{{\dot\beta}}
\def\dgamma{{\dot\gamma}}
\def\ddelta{{\dot\delta}}
\def\Sbar{\bar S}
\def\Im{{\rm Im}}
\def\sst{\scriptscriptstyle}
\def\cld{C_{\sst\rm LD}^{}}
\def\csd{C_{\sst\rm SD}^{}}
\def\bigI{{\rm I}_{\sst 3\rm D}}
\def\Mr{{\rm M}_{\sst R}}
\def\cJ{C_{\sst J}}
\def\one{{\sst(1)}}
\def\two{{\sst(2)}}
\def\vsd{v^{\sst\rm SD}}
\def\vasd{v^{\sst\rm ASD}}
\def\Phibar{\bar\Phi}
\def\F{{\cal F}_{\sst\rm SW}}
\def\P{{\cal P}}
\def\A{{\cal A}}
\def\susy{supersymmetry}
\def\sigmabar{\bar\sigma}
\def\barsigma{\sigmabar}
\def\ASD{{\scriptscriptstyle\rm ASD}}
\def\cl{{\,\rm cl}}
\def\lambdabar{\bar\lambda}
\def\R{{R}}
\def\psibar{\bar\psi}
\def\sqrtwo{\sqrt{2}\,}
\def\etabar{\bar\eta}
\def\Thetabar{{\bar\Theta_0}}
\def\Qbar{\bar Q}
\def\susic{supersymmetric}
\def\vhiggs{{\rm v}}
\def\vhiggsa{{\cal A}_{\sst00}}
\def\vbarhiggs{\bar{\rm v}}
\def\vhiggsbar{\bar{\rm v}}
\def\novetal{Novikov et al.}
\def\Novetal{Novikov et al.}
\def\ADS{Affleck, Dine and Seiberg}
\def\ads{Affleck, Dine and Seiberg}
\def\setI{\{{\cal I}\}}
\def\Abar{A^\dagger}
\def\B{{\cal B}}
\def\infinity{\infty}
\def\C{{\cal C}}
\def\Psitwo{\Psi_{\scriptscriptstyle N=2}}
\def\Psibartwo{\bar\Psi_{\scriptscriptstyle N=2}}
\def\ms{Minkowski space}
\def\zero{{\scriptscriptstyle(0)}}
\def\new{{\scriptscriptstyle\rm new}}
\def\u{\underline}
\def\uA{\,\lower 1.2ex\hbox{$\sim$}\mkern-13.5mu A}
\def\uX{\,\lower 1.2ex\hbox{$\sim$}\mkern-13.5mu X}
\def\uD{\,\lower 1.2ex\hbox{$\sim$}\mkern-13.5mu {\rm D}}
\def\uDzero{{\uD}^\zero}
\def\uAzero{{\uA}^\zero}
\def\upsizero{{\upsi}^\zero}
\def\uF{\,\lower 1.2ex\hbox{$\sim$}\mkern-13.5mu F}
\def\uW{\,\lower 1.2ex\hbox{$\sim$}\mkern-13.5mu W}
\def\uWbar{\,\lower 1.2ex\hbox{$\sim$}\mkern-13.5mu {\overline W}}
\def\Dbar{D^\dagger}
\def\Fbar{F^\dagger}
\def\uAbar{{\uA}^\dagger}
\def\uAbarzero{{\uA}^{\dagger\zero}}
\def\uDbar{{\uD}^\dagger}
\def\uDbarzero{{\uD}^{\dagger\zero}}
\def\uFbar{{\uF}^\dagger}
\def\uFbarzero{{\uF}^{\dagger\zero}}
\def\uV{\,\lower 1.2ex\hbox{$\sim$}\mkern-13.5mu V}
\def\uZ{\,\lower 1.2ex\hbox{$\sim$}\mkern-13.5mu Z}
\def\uv{\lower 1.0ex\hbox{$\scriptstyle\sim$}\mkern-11.0mu v}
\def\uc{\lower 1.0ex\hbox{$\scriptstyle\sim$}\mkern-11.0mu c}
\def\uPsi{\,\lower 1.2ex\hbox{$\sim$}\mkern-13.5mu \Psi}
\def\uPhi{\,\lower 1.2ex\hbox{$\sim$}\mkern-13.5mu \Phi}
\def\uchi{\lower 1.5ex\hbox{$\sim$}\mkern-13.5mu \chi}
\def\utheta{\lower 1.5ex\hbox{$\sim$}\mkern-13.5mu \theta}
\def\chitilde{\tilde \chi}
\def\etatilde{\tilde \eta}
\def\uchitilde{\lower 1.5ex\hbox{$\sim$}\mkern-13.5mu \tilde\chi}
\def\ueta{\lower 1.5ex\hbox{$\sim$}\mkern-13.5mu \eta}
\def\uetatilde{\lower 1.5ex\hbox{$\sim$}\mkern-13.5mu \tilde\eta}
\def\Psibar{\bar\Psi}
\def\uPsibar{\,\lower 1.2ex\hbox{$\sim$}\mkern-13.5mu \Psibar}
\def\upsi{\,\lower 1.5ex\hbox{$\sim$}\mkern-13.5mu \psi}
\def\uphi{\lower 1.5ex\hbox{$\sim$}\mkern-13.5mu \phi}
\def\psibar{\bar\psi}
\def\upsibar{\,\lower 1.5ex\hbox{$\sim$}\mkern-13.5mu \psibar}
\def\etabar{\bar\eta}
\def\uetabar{\,\lower 1.5ex\hbox{$\sim$}\mkern-13.5mu \etabar}
\def\chibar{\bar\chi}
\def\uchibar{\,\lower 1.5ex\hbox{$\sim$}\mkern-13.5mu \chibar}
\def\upsibarzero{\,\lower 1.5ex\hbox{$\sim$}\mkern-13.5mu \psibar^\zero}
\def\ulambda{\,\lower 1.2ex\hbox{$\sim$}\mkern-13.5mu \lambda}
\def\ulambdabar{\,\lower 1.2ex\hbox{$\sim$}\mkern-13.5mu \lambdabar}
\def\ulambdabarzero{\,\lower 1.2ex\hbox{$\sim$}\mkern-13.5mu \lambdabar^\zero}
\def\ulambdabarnew{\,\lower 1.2ex\hbox{$\sim$}\mkern-13.5mu \lambdabar^\new}
\def\D{{\cal D}}
\def\M{{\cal M}}
\def\N{{\cal N}}
\def\Dslash{\,\,{\raise.15ex\hbox{/}\mkern-12mu \D}}
\def\Dbarslash{\,\,{\raise.15ex\hbox{/}\mkern-12mu {\bar\D}}}
\def\delslash{\,\,{\raise.15ex\hbox{/}\mkern-9mu \partial}}
\def\delbarslash{\,\,{\raise.15ex\hbox{/}\mkern-9mu {\bar\partial}}}
\def\L{{\cal L}}
\def\hf{{\textstyle{1\over2}}}
\def\quarter{{\textstyle{1\over4}}}
\def\twe{{\textstyle{1\over12}}}
\def\eighth{{\textstyle{1\over8}}}
\def\fourth{\quarter}
\def\wb{Wess and Bagger}
\def\xibar{\bar\xi}
\def\ss{{\scriptscriptstyle\rm ss}}
\def\sc{{\scriptscriptstyle\rm sc}}
\def\uvcl{{\uv}^\cl}
\def\uAcl{\,\lower 1.2ex\hbox{$\sim$}\mkern-13.5mu A^{}_{\cl}}
\def\uAbarcl{\,\lower 1.2ex\hbox{$\sim$}\mkern-13.5mu A_{\cl}^\dagger}
\def\upsinew{{\upsi}^\new}
\def\ASDzero{{{\scriptscriptstyle\rm ASD}\zero}}
\def\SDzero{{{\scriptscriptstyle\rm SD}\zero}}
\def\SD{{\scriptscriptstyle\rm SD}}
\def\varthetabar{{\bar\vartheta}}
\def\three{{\scriptscriptstyle(3)}}
\def\dagthree{{\dagger\scriptscriptstyle(3)}}
\def\ld{{\scriptscriptstyle\rm LD}}
\def\vld{v^\ld}
\def\Dld{{\rm D}^\ld}
\def\Fld{F^\ld}
\def\Ald{A^\ld}
\def\Fbarld{F^{\dagger\scriptscriptstyle\rm LD}}
\def\Abarld{A^{\dagger\scriptscriptstyle \rm LD}}
\def\lambdald{\lambda^\ld}
\def\lambdabarld{\bar\lambda^\ld}
\def\psild{\psi^\ld}
\def\psibarld{\bar\psi^\ld}
\def\dsiginst{d\sigma_{\scriptscriptstyle\rm inst}}
\def\xione{\xi_1}
\def\xionebar{\bar\xi_1}
\def\xitwo{\xi_2}
\def\xitwobar{\bar\xi_2}
\def\thetatwo{\vartheta_2}
\def\thetatwobar{\bar\vartheta_2}
\def\Ltwo{\L_{\sst SU(2)}}
\def\Leff{\L_{\rm eff}}
\def\Laux{\L_{\rm aux}}
\def\oneloop{{\sst\rm 1\hbox{-}\sst\rm loop}}
\def\LSUtwo{{\cal L}_{\rm SU(2)}}
\def\Dhat{\hat\D}
\def\bkgd{{\sst\rm bkgd}}
\def\Lgft{{\cal L}_{\sst\rm g.f.t.}}
\def\Lghost{{\cal L}_{\sst\rm ghost}}
\def\Sinst{S_{\rm inst}}
\def\etal{{\rm et al.}}
\def\S{{\cal S}}

\newcommand{\nd}[1]{/\hspace{-0.5em} #1}
\begin{titlepage}
\begin{flushright}
NI-97019 \\
DTP-97/20 \\
March 1997 \\
hep-th/9703228
\end{flushright}
\begin{centering}
\vspace{.2in}
{\large {\bf Instantons, Three-Dimensional Gauge Theory, \\
and the Atiyah-Hitchin Manifold}}\\
\vspace{.4in}
 N. Dorey$^{1}$, V. V. Khoze$^{2}$, M. P. Mattis$^{3}$, D. Tong$^{1}$  and 
S. Vandoren$^{1}$\\
\vspace{.4in}
$^{1}$ Physics Department, University of Wales Swansea \\
Singleton Park, Swansea, SA2 8PP, UK\\
\vspace{.3in}
$^{2}$ Physics Department, Centre for Particle Theory  \\
University of Durham, Durham DH1 3LE, UK \\
\vspace{.3in} 
$^{3}$ Theoretical Division,  Los Alamos National Laboratory \\
Los Alamos, NM 87545, USA\\
\vspace{.4in}
{\bf Abstract} \\
\end{centering}
We investigate quantum effects on the Coulomb 
branch of three-dimensional $N=4$ supersymmetric gauge theory with gauge 
group $SU(2)$. We calculate perturbative and one-instanton contributions 
to the Wilsonian effective action using standard weak-coupling methods. 
Unlike the four-dimensional case, and despite supersymmetry, the contribution 
of non-zero modes to the instanton measure does not cancel. Our results 
allow us to fix the weak-coupling boundary conditions for the differential 
equations which determine the hyper-K\"{a}hler metric on the quantum moduli 
space. We confirm the proposal of Seiberg and Witten that the Coulomb branch 
is equivalent, as a hyper-K\"{a}hler manifold, to the centered moduli space 
of two BPS monopoles constructed by Atiyah and Hitchin.


\end{titlepage}
\section{Introduction}
\paragraph{}
Recent work by several authors \cite{seib}-\cite{ber2} 
has provided exact information about the low-energy dynamics of 
$N=4$ supersymmetric gauge theory in three dimensions. In particular, 
an interesting connection has emerged between the quantum 
moduli spaces of these theories and the classical moduli spaces of BPS
monopoles in $SU(2)$ gauge theory \cite{SW3}. 
Chalmers and Hanany \cite{chahan} have proposed that the Coulomb branch of
the $SU(n)$ gauge theory is equivalent as a hyper-K\"{a}hler manifold 
to the centered moduli space of $n$ BPS monopoles. For $n>2$ this
correspondence is intriguing because the hyper-K\"{a}hler
metric on the manifold in question is essentially unknown. 
Subsequently, Hanany and Witten \cite{hanwit} have 
shown that the equivalence, for all $n$, is a consequence of 
S-duality applied to a certain configuration of D-branes in 
type IIB superstring theory. 
\paragraph{}
The case $n=2$, which is the main topic of this paper, 
provides an important test for these ideas because the
two-monopole moduli space and its hyper-K\"{a}hler metric have 
been found explicitly by Atiyah and Hitchin \cite{AH}. We will refer to the
four-dimensional manifold which describes the relative separation and
charge angle of two BPS monopoles as the Atiyah-Hitchin (AH)
manifold. In fact these authors
effectively classified all four-dimensional 
hyper-K\"{a}hler manifolds with an $SO(3)$
action which rotates the three inequivalent complex structures. 
Correspondingly, the moduli space of the 
$N=4$ theory with gauge group $SU(2)$ was
analysed by Seiberg and Witten \cite{SW3}. 
The effective low-energy theory in this case is a non-linear $\sigma$-model 
with the four-dimensional Coulomb branch of the $SU(2)$ theory 
as the target manifold.  
The $N=4$ supersymmetry of the low-energy theory requires that the 
metric induced on the target space by the $\sigma$-model 
kinetic terms be hyper-K\"{a}hler \cite{ag}.  
By virtue of its global symmetry structure,
the Coulomb branch of this theory necessarily fits into Atiyah and Hitchin's
classification scheme. Seiberg and Witten compared the weak coupling
behaviour of the SUSY gauge theory with the asymptotic form of the
metric on the AH manifold in the limit of large-spatial separation
between the monopoles, $r\gg 1$ \footnote{$r$ is the separation between
the monopoles in units of the inverse gauge-boson mass in the $3+1$
dimensional gauge theory in which the two BPS monopoles live.}. 
They found exact agreement between 
perturbative effects in the SUSY gauge theory, and the expansion 
of the AH metric in inverse powers of $r$. 
In fact, as we will review below, there are an infinite
number of inequivalent hyper-K\"{a}hler four-manifolds with the required
isometries which share this asymptotic behaviour. However, Atiyah and
Hitchin showed that only one of these, the AH manifold itself, is 
singularity-free. Motivated by expectations from string
theory \cite{seib}, Seiberg and Witten proposed that the Coulomb branch 
of the three-dimensional 
$N=4$ theory should also have no singularities. The correspondence
between the two manifolds then follows automatically.   
\paragraph{}
Clearly the arguments reviewed above come close to a 
first-principles demonstration that the Coulomb branch of the
$SU(2)$ theory is the AH manifold. The only assumption made about the
strong coupling behaviour of the SUSY gauge theory which is not
automatically guaranteed by symmetries alone is the absence of
singularities. In this paper we will proceed 
without this assumption\footnote{In the
following, we will, however, retain the
weaker assumption that the moduli space has at most isolated
singularities.}. One must then choose
between an infinite number of possible metrics with the same 
asymptotic form. 
By virtue of the hyper-K\"{a}hler condition and
the global symmetries, the components of these metrics each satisfy the
same set of coupled non-linear ODE's as the AH metric. 
In fact, as we review below, 
there is precisely a one-parameter family of solutions of these
differential equations which have the same asymptotic 
behaviour as the AH metric to all finite orders in $1/r$ but differ by terms
of order $\exp(-r)$. Seiberg and Witten showed that
the exponentially suppressed 
terms correspond to instanton effects in the weakly-coupled 
SUSY gauge theory. In this paper we will calculate the one-loop perturbative 
and one-instanton contributions 
to the low-energy theory using standard background field and semiclassical
methods respectively. The results of these calculations suffice to 
fix the boundary 
conditions for the differential equations which determine the metric. 
We find that the resulting metric is equal to the AH metric, thereby
confirming the prediction of Seiberg and Witten. 
\paragraph{}
The paper is organised as follows. 
In Section 2, we introduce the model
and review its relation to $N=2$ SUSY Yang-Mills theory in
four dimensions. We also review the classical form of the Coulomb
branch and perform an explicit one-loop evaluation of the metric. 
In Section 3 we discuss the properties of supersymmetric instantons in
three-dimensional (3D) gauge theory. The field configurations in question
themselves correspond to the BPS monopoles of the four-dimensional (4D)
gauge theory\footnote{To avoid confusion with the proposed connection
to monopole moduli spaces, from now on, the term
`monopole' will always refer to the instantons of the three-dimensional
SUSY gauge theory unless otherwise stated.}. Much of Section 3 is devoted to 
obtaining the measure for integration over the 
instanton collective coordinates. The number of bosonic and fermionic zero
modes of the BPS monopole is determined by the Callias index theorem
which we briefly review. Like their four-dimensional counterparts, the
three-dimensional instantons have a self-duality property which,
together with supersymmetry, ensures a large degree of cancellation
between non-zero modes of the bose and fermi fields. However, 
unlike the four-dimensional case and despite supersymmetry, 
this cancellation is not
complete because of the spectral asymmetry of the Dirac operator in a
monopole background. We calculate, for the first time, 
the residual term which arises from
the non-cancelling ratio of determinants of the quadratic fluctuation
operators of the scalar, fermion, gauge and ghost degrees of freedom. 
We apply the resulting one-instanton measure to calculate the
leading non-perturbative correction to a four-fermion vertex in the
low-energy effective action. 
\paragraph{}
In Section 4 we show that the one-loop and one-instanton data calculated 
in Sections 2 and 3, together with the (super-)symmetries of the model,  
are sufficient to to determine the exact metric on the Coulomb branch.    
We begin by reviewing the arguments leading to the exact solution of
the low-energy theory proposed by Seiberg and Witten. We analyze the
solutions of the non-linear ODE's which determine the metric and exhibit a
one-parameter family of solutions which agree with the metric on the
Coulomb branch determined up to one-loop in perturbation theory. We show
that each of these solutions leads to a different prediction for the
one-instanton effect calculated in Section 3 and precise agreement is
obtained only for the solution which corresponds to the
AH-manifold: the singularity-free case. For the most part, calculational 
details are relegated to a series of Appendices.      

\section{$N=4$ Supersymmetry in 3D}
\subsection{Fields, symmetries and dimensional reduction} 
\paragraph{}
In this Section we will briefly review some basic facts about the
$N=4$ supersymmetric $SU(2)$ gauge theory in three dimensions considered in
\cite{SW3}. It is particularly convenient to obtain this theory from
the four-dimensional Euclidean $N=2$ SUSY Yang-Mills theory by dimensional
reduction. In discussing the 4D theory we will adopt the notation
and conventions of \cite{dkm1}: the $N=2$ gauge multiplet contains the
gauge field $\uv_{m}$ ($m=0,1,2,3$), a complex scalar $\uA$ and two species of Weyl
fermion $\ulambda_{\alpha}$ and $\upsi_{\alpha}$ all in the adjoint
representation of the gauge group. As in \cite{dkm1} we use undertwiddling 
for the fields  in the $SU(2)$ matrix notation,
$\uX \equiv X^a\tau^a/2$. The resulting 
$N=2$ SUSY algebra admits an $SU(2)_{\cal{R}}\times
U(1)_{\cal{R}}$ group of automorphisms.  In the 4D 
quantum theory, the Abelian factor of the 
${\cal R}$-symmetry group is anomalous due to the effect of 
4D instantons. 
\paragraph{}
Following Seiberg and Witten, the three-dimensional theory is obtained
by compactifying one spatial dimension\footnote{
For simplicity of presentation we choose $x_{3}$ 
to be the compactified dimension. In practice, however, in order
to flow from the standard chiral basis of gamma matrices in 4D 
to gamma matrices in 3D 
one has to dimensionally reduce in the
$x_2$ direction. To remedy this one can always reshuffle gamma matrices
in 4D. See Appendix A for more details.}, say $x_{3}$, 
on a circle of radius $R$. 
In the following we will restrict our attention 
to field configurations which are 
independent of the compactified dimension. This yields a classical
field theory in three spacetime dimensions
which we will then quantize. 
Seiberg and Witten also consider the distinct problem 
of quantizing the $N=2$ theory on $R^{3}\times S^{1}$. 
In this approach quantum fluctuations of the fields which depend on $x_{3}$
are included in the path integral and one can interpolate between the
3D and 4D quantum theories by varying $R$. We will not consider
this more challenging problem here. 
Integrating over $x_{3}$ in the action gives, 
\begin{eqnarray}
\frac{1}{g^{2}}\int\, d^{4}x & \rightarrow &  \frac{2\pi}{e^{2}}\int\, d^{3}x
\label{coupling}
\end{eqnarray}
where $e=g/\sqrt{R}$ defines the dimensionful 3D gauge coupling in
terms of the dimensionless 4D counterpart $g$. 
\paragraph{}
Compactifying one dimension breaks the $SO(4)_{E}\simeq
SU(2)_{l}\times SU(2)_{r}$ group of rotations of 
four-dimensional Euclidean spacetime down to $SO(3)_{E}$. Following
the notation of \cite{SW3}, the
double-cover of the latter group is denoted $SU(2)_{E}$. 
The 4D gauge
field $\uv_{m}$ splits into a 3D gauge field $\uv_{\mu}$, 
and a real scalar $\uphi_{3}$ such that 
$ \uv_{\mu}=\uv_{m}$, for $m=\mu=0,1,2$ and $\uphi_{3}=\uv_{3} $.
It
is also convenient to decompose the 4D complex scalar $\uA$ into two real 
scalars: $\uphi_{1}=\sqrt{2}{\rm Re}\uA$  and $\uphi_{2}=\sqrt{2}{\rm 
Im}\uA$. The 4D Weyl spinors $\ulambda_{\alpha}$, $\upsi_{\alpha}$ 
($\ulambdabar_{\dot{\alpha}}$, $\upsibar_{\dot{\alpha}}$) of
$SU(2)_{l}$ ($SU(2)_{r}$) can be rearranged to form four 3D Majorana
spinors of $SU(2)_{E}$: $\uchi^{A}_{\alpha}$ for $A=1,2,3,4$. 
Correspondingly, the two Weyl supercharges of the $N=2$ theory are
reassembled as four Majorana supercharges which generate the $N=4$ 
supersymmetry of the three-dimensional theory. Details of dimensional 
reduction, field definitions and our 
conventions for spinors in three and four dimensions 
are given in Appendix A. 
\paragraph{}
While the number of spacetime symmetries decreases reducing from 4D
down to 3D, the number of ${\cal R}$-symmetries increases. First, 
the $SU(2)_{{\cal R}}$ symmetry which rotates the two species of Weyl
fermions and supercharges in the four-dimensional theory remains
unbroken in three dimensions. In addition, the $U(1)_{\cal{R}}$
symmetry is enlarged to a simple group which, following \cite{SW3}, 
we will call $SU(2)_{N}$. Unlike $U(1)_{\cal R}$ in the 4D case, this
symmetry remains unbroken at the quantum level. 
The real scalars, $\uphi_{i}$ with $i=1,2,3$,
transform as a ${\bf 3}$ of $SU(2)_{N}$. 
The index $A=1,2,3,4$ on the Majorana spinors $\uchi^{A}_{\alpha}$ introduced
above reflects the 
fact that they transform as a ${\bf 4}$ of the combined ${\cal
R}$-symmetry group, $SO(4)_{\cal R}\simeq SU(2)_{\cal R}\times
SU(2)_{N}$. 
\subsection{The low-energy theory}
\paragraph{}          
The three-dimensional $N=4$ theory has flat directions along which the
three adjoint scalars $\uphi_{i}$ acquire mutually commuting 
expectation values. By a gauge
rotation, each of the scalars can be chosen to lie in the third isospin
component: $\langle \uphi_{i} \rangle
=\sqrt{2}{\rm v}_{i}\tau^{3}/2$. After modding out the action of the Weyl
group, ${\rm v}_{i}\rightarrow -{\rm v}_{i}$, the three real parameters 
${\rm v}_{i}$
describe a manifold of gauge-inequivalent vacua. As we will see below,
this manifold is only part of the 
classical Coulomb branch. For any non-zero
value of ${\bf v}=({\rm v}_{1},{\rm v}_{2},{\rm v}_{3})$, 
the gauge group is broken down to $U(1)$  and two of the gauge bosons
acquire masses $M_{W}=\sqrt{2}|{\bf v}|$ by the adjoint Higgs
mechanism. At the classical level,  the global $SU(2)_{N}$
symmetry is also 
spontaneously broken to an Abelian subgroup $U(1)_{N}$ on the Coulomb
branch. The remaining component of the gauge field is the massless
photon ${v}_{\mu}={\rm Tr}(\uv_{\mu}\tau_{3})$ with Abelian
field-strength $v_{\mu \nu}$. 
For each matrix-valued field $\uX$ in the
microscopic theory, we define a corresponding massless field in the
Abelian low-energy theory: $X={\rm Tr}(\uX\tau^{3})$.  Hence, at
the classical level, the
bosonic 
part of the low-energy Euclidean action is simply given by the free massless 
expression,    
\begin{equation}
S_{\rm B}=\frac{2\pi}{e^{2}}
\int\,d^{3}x \left[
\quarter v_{\mu \nu}v_{\mu \nu}+\hf\partial_{\mu}\phi_{i}
\partial_{\mu}\phi_{i}\right]
\end{equation}
\paragraph{}
The presence of 3D instantons in the theory means 
we must also include a surface term
in the action, which is analogous to the $\theta$-term in four dimensions. 
In the low-energy theory this term can be written as, 
\begin{equation}
S_{\rm S}=\frac{i\sigma}{8\pi}\int\, d^{3}x\, 
\varepsilon^{\mu\nu\rho}\partial_{\mu}v_{\nu\rho} 
\label{surface}
\end{equation}
A dual description of the low-energy theory can be obtained by
promoting the parameter $\sigma$ to be a dynamical field \cite{pol}. 
This field serves as a Lagrange multiplier for the Bianchi identity
constraint. In the presence of this constraint one may integrate out
the Abelian field strength to obtain the bosonic effective action,
\begin{equation}
S_{\rm B}=\frac{2\pi}{e^{2}}
\int\,d^{3}x \ \hf\partial_{\mu}\phi_{i}
\partial_{\mu}\phi_{i} +\frac{2e^{2}}{\pi(8\pi)^{2}} \int\, d^{3}x \ 
\hf\partial_{\mu}\sigma\partial_{\mu}\sigma   
\label{lcl}
\end{equation}
The Dirac quantization of magnetic charge, 
or, equivalently, the 3D instanton
topological charge,
\begin{equation}
k \ = \ \frac{1}{8\pi}\int\, d^{3}x\, 
\varepsilon^{\mu\nu\rho}\partial_{\mu}v_{\nu\rho} \ \in  Z \ ,
\label{tc3d}
\end{equation}
means that, with the
normalization given in (\ref{surface}), $\sigma$ is a periodic
variable with period $2\pi$. In the absence of magnetic charge
$\sigma$ only enters through its derivatives and the action has a
trivial symmetry, $\sigma \rightarrow \sigma+ c$ where $c$ is a
constant. The VEV of the $\sigma$-field spontanteously breaks this symmetry
and provides an extra compact dimension for the Coulomb
branch. Modding out the action of the Weyl group, the
classical Coulomb branch 
can then be thought of as $(R^{3}\times S^{1})/Z_{2}$. 
\paragraph{}
It will be convenient to write the 
fermionic terms in the low-energy action in terms of the (dimensionally
reduced) Weyl fermions. At the classical level,
the action contains free kinetic terms for these massless degrees of freedom, 
\begin{equation}
S_{\rm F}=\frac{2\pi}{e^{2}}
\int\,d^{3}x \, \left(i\bar{\lambda}\bar{\sigma}_{\mu}\partial_{\mu}\lambda+      i\bar{\psi}\bar{\sigma}_{\mu}\partial_{\mu}\psi \right)
\label{lclf}
\end{equation}
The Weyl fermions in 4D
can be related to the 3D Majorana 
fermions $\chi^{A}_{\alpha}$ by going to a complex basis for the 
$SO(4)_{{\cal R}}$ index. In Appendix A we define a basis such that the 
holomorphic components $\chi^{a}_{\alpha}$ are equal to 
$\epsilon_{\alpha\dot{\beta}}\bar{\lambda}^{\dot{\beta}}$ and 
$\epsilon_{\alpha\dot{\beta}}\bar{\psi}^{\dot{\beta}}$ for $a=1$ and $a=2$ 
respectively. Similarly the anti-holomorphic components 
$\chi^{\bar{a}}_{\alpha}$ are equal to the left-handed Weyl fermions 
$\lambda_{\alpha}$ and $\psi_{\alpha}$ for $\bar{a}=\bar{1}$ and 
$\bar{a}=\bar{2}$. In this basis, the effective action $(\ref{lclf})$ 
can be rewritten as,
\begin{equation}
S_{\rm F}=-\frac{2\pi}{e^{2}}
\int\,d^{3}x \, 
\delta_{a\bar{b}}\chi^{a}\gamma_{\mu}\partial_{\mu}\chi^{\bar{b}}      
\label{lclf2}    
\end{equation}
where  $\gamma_\mu$ are gamma-matrices in 3D.
\paragraph{}
Perturbative corrections lead to finite corrections to the
classical low-energy theory in powers of $e^{2}/M_{W}$. At 
the one-loop level several effects occur. First, there is a finite
renormalization of the gauge coupling appearing in (\ref{lcl}) and
(\ref{lclf}): 
\begin{eqnarray}
\frac{2\pi}{e^{2}} & \rightarrow & \frac{2\pi}{e^{2}} -\frac{1}{2\pi
M_{W}} 
\label{1loop}
\end{eqnarray}
This result is demonstrated explicitly in Appendix B.
Second, there is a
more subtle one-loop effect discussed in \cite{SW3}. By
considering the realization of the $U(1)_{N}$ symmetry in an instanton
background, Seiberg and Witten showed that a particular coupling 
between the dual photon $\sigma$ and the other scalars $\phi_{i}$
appears in the one-loop effective action. More generally the 
one-loop effective action will contain vertices with arbitrary numbers of 
boson and fermion legs. However, as we will see in Section 4, the exact 
form of the low-energy effective action is essentially determined 
to all orders in perturbation theory by the 
finite shift in the coupling (\ref{1loop}) together with the constraints 
imposed by $N=4$ supersymmetry. In the next Section we will turn our 
attention to the non-perturbative effects which modify this 
description.        
     
\section{Supersymmetric Instantons in Three Dimensions}
\subsection{Instantons in the microscopic theory} 
\paragraph{}   
The four-dimensional $N=2$ theory has static BPS monopole solutions of 
finite energy \cite{bog}, \cite {PS}. 
In the three-dimensional theory obtained by
dimensional reduction, these field configurations have finite Euclidean
action, $S_{\rm cl}^{(k)}=|k|S_{\rm cl}$ for magnetic charge $k$, with 
$S_{\rm cl}=(8\pi^{2}M_{W})/e^{2}$. For each value of $k$, 
the monopole solutions are exact minima of the action which yield 
contributions of order $\exp(-|k|S_{\rm cl})$ 
to the partition function and Greens functions of the theory at weak
coupling. Hence BPS monopoles appear as instantons in the $N=4$
supersymmetric gauge theory in 3D. In this Section 
(together with Appendix C), as well as presenting several general results, we 
will provide a quantitative analysis of 
these effects for the case $k=1$. We begin by determining the bosonic and 
fermionic zero modes of the instanton and the corresponding 
collective coordinates. We then consider the contribution of 
non-zero frequency modes to the instanton measure which requires the 
evaluation of a ratio of functional determinants. 
In subsection 3.2, we use these results 
to calculate the leading non-perturbative correction to a 
four-fermion vertex in the low-energy effective action.    
\paragraph{}
To exhibit the properties of the 3D instantons, 
it is particularly convenient to work with the (dimensionally-reduced)
fields of the four-dimensional theory and also with the specific
vacuum choice ${\rm v}_{i}={\rm v}\delta_{i3}$. 
In this case, 
the static Bogomol'nyi 
equation satisfied by the gauge and Higgs components of the monopole 
can be concisely rewritten as a self-dual
Yang-Mills equation for the four-dimensional gauge field, $\uv_{m}$ of 
Section 2.1 
\cite{lohe}\footnote{Of course these is no loss of generality here, 
for any choice of 
three-dimensional vacuum ${\rm v}_{i}$ 
it is possible to construct an analogous 
four-dimensional Euclidean gauge field which is
self-dual in the monopole background: one simply needs to include the
component ${\rm v}_{i}\phi_{i}$ of the scalar as the `fourth' component of
the gauge field. However, only if we choose ${\rm v}_{i}={\rm v}\delta_{i3}$ 
does this correspond to the four-dimensional gauge field of Section 2.1.}   
\begin{equation}
\uv^{\rm cl}_{mn}=^{*}\uv^{\rm cl}_{mn}  
\label{SD}
\end{equation}
Because of this self-duality, instantons in three dimensions have many
features in common with their four-dimensional counterparts. In the 
following, we will focus primarily on the new features which are
special to instanton effects in three-dimensional gauge theories. One
important difference is that instantons in 3D are exact solutions
of the equations of motion in the spontaneously broken phase. In four
dimensions, instantons are only quasi-solutions in the presence of
a VEV and this leads to several complications which do not occur in the 
3D case.   
\paragraph{}
Bosonic zero modes, $\delta\uv^{m}=\uZ^{m}$, 
of the 3D $k$-instanton configuration are 
obtained by solving the 
linearized self-duality equation subject to the background field gauge
constraint (see Appendix C),
\begin{equation}
{\cal D}_{\rm cl}^{[m}{\uZ}^{n]}=\,^{*}{\cal D}_{\rm cl}^{[m}{\uZ}^{n]}
\ \ \ ,\ \ \ {\cal D}_{{\rm cl}}^{m}{\uZ}^{m}=0 \ .
\label{fan}
\end{equation}
where, ${\cal D}^{m}_{\rm cl}$ is the adjoint gauge-covariant derivative 
in the self-dual gauge background, $\uv^{\rm cl}$.   
As a consequence of self-duality, there
is an exact correspondence between these bosonic zero modes and the 
fermionic zero modes, which are solutions of the adjoint Dirac equation in the
self-dual background. The latter is precisely the equation of motion for the 
4D Weyl fermions in the monopole background, 
\begin{eqnarray}
{\ssl \bar{\cal D}_{\rm cl}^{\dot{\alpha}\alpha}}\ulambda^{\rm cl}_{\alpha} & =
& 0 \label{dbar} \\
{\ssl {\cal D}}_{\rm cl}^{\alpha\dot{\alpha}}
\bar{\ulambda}^{\rm cl}_{\dot{\alpha}} & =
& 0 
\label{d}
\end{eqnarray}
Following Weinberg \cite{w1}, we form
the bispinor operators, 
$\Delta_{+}=\ssl{\bar{\cal D}_{\rm cl}}\ssl{\cal D}_{\rm cl}$ and 
$\Delta_{-}=\ssl{{\cal D}}_{\rm cl}\ssl{\bar{\cal D}}_{\rm cl}$.
In general, for a
self-dual background field\footnote{We define standard 
self-dual and anti-self-dual projectors,
$\sigma^{mn}=\quarter\sigma^{[m}\sigmabar^{n]}$ and    
$\sigmabar^{mn}=\quarter\sigmabar^{[m}\sigma^{n]}$.
}, $\uv^{\rm cl}_m$, 
\begin{eqnarray}
\left(\Delta_{+}\right)_{\dot{\alpha}}^{\ \dot{\beta}} & = &   
{\cal D}_{\rm cl}^{2}\delta_{\dot{\alpha}}^{\ \dot{\beta}} +
\left(\sigmabar^{mn}\right)_{\dot{\alpha}}^{\ \dot{\beta}}\uv^{\rm cl}_{mn}
 =
{\cal D}_{\rm cl}^{2}\delta_{\dot{\alpha}}^{\ \dot{\beta}} \ , 
\nonumber \\ \left(\Delta_{-}\right)_{\alpha}^{\ \beta} & = &  
{\cal D}_{\rm cl}^{2}\delta_{\alpha}^{\ \beta} +\left(\sigma^{mn}\right)
_{\alpha}^{\ \beta}\uv^{\rm cl}_{mn}
\label{ddbar}
\end{eqnarray}
$\Delta_{-}$ can have normalizable zero modes, while $\Delta_{+}$ 
is positive and has none. Let the number of normalizable 
zero modes of $\Delta_{-}$ be $q$.
Then correctly accounting for spinor indices, the number of normalizable 
zero modes, 
${\uZ}_m$, of the gauge
field $\uv^{\rm cl}_{m}$ is $2q$ \cite{w1}. 
\paragraph{}
Adapting the Callias index theorem \cite{cal} to the
context of BPS monopoles, Weinberg \cite{w1} 
showed that $q$ could be obtained as
the $\mu \rightarrow 0$ limit of the regularized trace, 
\begin{equation}  
{\cal I}(\mu)={\rm Tr}\left[\frac{\mu}{\Delta_{-}+\mu} -
\frac{\mu}{\Delta_{+}+\mu}\right]
\label{Callias}
\end{equation}
defined for $\mu>0$. It turns out that the only non-zero contribution to
${\cal I}(\mu)$ comes from a surface term that can be evaluated
explicitly for arbitrary $k$. Weinberg's result is, 
\begin{equation}
{\cal I}(\mu)=\frac{2kM_{W}}{(M_{W}^{2}+\mu)^{\frac{1}{2}}}
\label{Weinberg}
\end{equation}
Setting $\mu=0$ yields $q=2k$. Another consequence of this analysis is
that the adjoint Dirac equation (\ref{dbar}) has $2k$ independent
solutions while (\ref{d}) has none. Although the index information is
contained in the $\mu\rightarrow 0$ limit of equation (\ref{Weinberg}) we will
need this result for general $\mu$ in the following. 
\paragraph{}
The upshot of Weinberg's index calculation is the conventional
wisdom that the BPS monopole of charge $k$,
or, equivalently, the 3D $k$-instanton, has $4k$ bosonic
collective coordinates. 
For $k=1$ these simply correspond to the three components, $X_{m}$, 
of the instanton position in three-dimensional spacetime and an additional
angle, $\theta \in [0, 2\pi]$, which describes the orientation of the 
instanton in the unbroken $U(1)$ gauge subgroup. In an instanton
calculation we must integrate over these coordinates with the measure
obtained by changing variables in the path integral. Explicitly, 
\begin{equation}
\int \, d\mu_{B}=\int\,\frac{d^{3}X}{(2\pi)^{\frac{3}{2}}} 
({\cal J}^{}_{X})^{\frac{3}{2}} \int_{0}^{2\pi}\,\frac{d\theta}
{(2\pi)^{\frac{1}{2}}} ({\cal J}_{\theta})^{\frac{1}{2}}
\label{bmeasure}
\end{equation}
In Appendix C, we calculate the Jacobian factors, 
${\cal J}^{}_{X}=S_{\rm cl}$ and ${\cal J}_{\theta}=S_{\rm cl}/M_{W}^{2}$. 
\paragraph{}
Similarly the two species of Weyl fermions, $\ulambda$ and $\upsi$ each have 
$2k$ independent zero-mode solutions in the instanton-number $k$
background. For $k=1$ these four modes correspond
to the action of the four supersymmetry generators under which the
3D instanton transforms non-trivially. As in the four-dimensional case, 
the modes in question can be parametrized in terms of two-component
Grassmann collective coordinates $\xi_{\alpha}$ and $\xi'_{\alpha}$ as 
\begin{eqnarray}
\ulambda^{\rm cl}_{\alpha} & =& \hf\xi_{\beta}
(\sigma^{m}\sigmabar^{n})_{\alpha}^{\ \beta}
\uv^{\rm cl}_{mn} \nonumber \\ 
\upsi^{\rm cl}_{\alpha} & =& \hf\xi'_{\beta}
(\sigma^{m}\sigmabar^{n})_{\alpha}^{\ \beta}
\uv^{\rm cl}_{mn} 
\label{fmodes}
\end{eqnarray}
The corresponding contribution to the instanton measure is, 
\begin{equation}
\int \, d\mu_{F}=\int\, d^{2}\xi\, d^{2}\xi' ({\cal J}_{\xi})^{-2}
\label{fmeasure}
\end{equation}
In Appendix C we find that ${\cal J}_{\xi}=2S_{\rm cl}$.
\paragraph{}  
As usual in any saddle-point calculation, to obtain the leading-order
semiclassical result it is necessary to perform Gaussian integrals over
the small fluctuations of the fields around the classical 
background. In general these integrals yield functional determinants
of the operators which appear at quadratic order in the expansion
around the instanton. A simplifying feature that holds for all
self-dual configurations is that each of the fluctuation operators for
scalars, spinors gauge fields and ghosts 
are related in a simple way to one of the operators $\Delta_{+}$
or $\Delta_{-}$. This standard connection is as follows.
In the chiral basis for 4D Majorana fermions (see Appendix A), the 
quadratic fluctuation operator is 
\begin{equation}
\Delta_{F}=\left(\begin{array}{cc} 0 & \ssl {\cal D}_{\rm cl} 
\\ \ssl \bar{\cal D}_{\rm cl} & 
0 \end{array}\right)
\label{fermop}
\end{equation}
Performing the Grassmann Gaussian integration over $\ulambda$ yields
\begin{equation}
\frac{{\rm Pf}(\Delta_{F})}{{\rm Pf}(\Delta^{(0)}_{F})}=\left(\frac{{\rm det'}
(\Delta_{F}^{2})}{{\rm det}(\Delta^{(0)2}_{F})}\right)^{\frac{1}{4}}=
\left(\frac{{\rm det'}(\Delta_{-}){\rm det}(\Delta_{+})}{{\rm det}(\Delta
^{(0)})^{2}}\right)^{\frac{1}{4}}
\label{fermdet}
\end{equation}
where ${\rm det'}$ denotes the removal of zero eigenvalues and the 
superscript $(0)$ denotes the 
fluctuation operator for the corresponding field in the vacuum sector.   
In particular we define the operator $\Delta^{(0)}= 
\ssl{\bar{\cal D}}^{(0)}{\ssl{\cal D}}^{(0)}=\ssl{{\cal D}}^{(0)}
{\ssl{\bar{\cal D}}}^{(0)}$, formed from the vacuum Dirac operators 
${\ssl{\cal D}}^{(0)}$ and ${\ssl \bar{\cal D}}^{(0)}$. Integrating out the $\upsi$ field yields a second factor equal to (\ref{fermdet}). 
For the bose fields we introduce the gauge fixing and ghost terms using 
the 4D background gauge, ${\cal D}_m^{\rm cl} \delta \uv_{m}^{a} = 0$,
and expand the action around the classical 
configuration. Quadratic fluctuation operators for the 
complex scalar, $\uA$, the gauge field, $\uv$, and the ghosts, $\uc$ and 
$\bar{\uc}$ are 
\begin{eqnarray}
 & \Delta_{A} = \Delta_{c} = {\cal D}_{\rm cl}^{2} 
\ = \ \frac{1}{2}{\rm Tr}(\Delta_{+})
 \nonumber \\ & \Delta_{v} =  
-{\cal D}_{\rm cl}^{2}\delta_{mn}+2\uv^{\rm cl}_{mn} \ = \ 
-\hf{\rm Tr}(\bar{\sigma}^{n}\Delta_{-}\sigma^{m})
\label{boseop}
\end{eqnarray}
where ${\rm Tr}$ is over the spinor indices. The bosonic Gaussian 
integrations now give, 
\begin{equation}
\left(\frac{{\rm det'}(\Delta_{v})}{{\rm det}(\Delta^{(0)}_{v})}\right)^{
-\frac{1}{2}}\left(\frac{{\rm det}(\Delta_{A})}{{\rm det}(\Delta^{(0)}_{A})}
\right)^{-1}\left(\frac{{\rm det}(\Delta_{c})}{{\rm det}(\Delta^{(0)}_{c})}
\right)^{+1} = \left(\frac{{\rm det'}(\Delta_{-})}{{\rm det}(\Delta^{(0)}
)}\right)^{-1}
\label{bosedet}
\end{equation}
Combining the fermion and boson contributions, we find that the 
total contribution of non-zero modes to the instanton measure is given by,  
\begin{equation}
R=\left[\frac{{\rm det}(\Delta_{+})}{{\rm det'}(\Delta_{-})}
\right]^{\frac{1}{2}}
\label{cancel}
\end{equation} 
\paragraph{} 
In any supersymmetric theory the total number of non-zero eigenvalues
of bose and fermi fields is precisely equal. This corresponds to the fact 
that,  for any self-dual background,  
the non-zero eigenvalues of $\Delta_{+}$ and $\Delta_{-}$ are
equal \cite{adda}. Naively, this suggests that the ratio $R$ is unity. 
As the spectra of these two operators contain a continuum of
scattering states in addition to normalizable bound states, this
assertion must be considered carefully. For the continuum
contributions to the determinants of $\Delta_{+}$ and $\Delta_{-}$ to
be equal, it is necessary not only that the continuous eigenvalues
have the same range, but that the density of these eigenvalues should
also be the same. Following the original approach of 't Hooft \cite{th} in four
dimensions, one can regulate the problem by putting the system
in a spherical box with fixed boundary conditions. In this case the
spectrum of scattering modes becomes discrete and the resulting
eigenvalues depend on the phase shifts of the scattering
eigenstates. In the limit where the box size goes to infinity, these
phase shifts determine the density of continuum eigenvalues. For a 
four-dimensional instanton, 
't Hooft famously discovered that the phase shifts in
question were equal for the small fluctuation operators of each of the
fields. As a direct result of this, the ratio $R$ is equal to unity in
the background of any number of instantons in a 4D supersymmetric
gauge theory. However, the phase-shifts associated with the operators 
$\Delta_{+}$ and $\Delta_{-}$  are not equal in a monopole
background. In the four-dimensional theory, 
Kaul \cite{k} noticed that this effect leads to a non-cancellation of 
quantum corrections to the monopole mass. In our case, as we will show
below, it will yield a non-trivial value of $R$. 
\paragraph{}
In fact the mismatch in the continuous spectra of $\Delta_{+}$ and
$\Delta_{-}$ can be seen already from the index function
${\cal I}(\mu)$. Comparing with the definition (\ref{Callias}), we see
that the fact that Weinberg's formula (\ref{Weinberg}) has a
non-trivial dependence on $\mu$ and is not simply equal to $2k$
precisely indicates a difference between the non-zero spectra of the two
operators. This observation can be made precise by the following
steps. First, dividing
(\ref{Callias}) by $\mu$ and then performing a parametric integration
we obtain.
\begin{equation}
\int_{\mu}^{\infty}\,\frac{d\mu'}{\mu'}\, {\cal I}(\mu')={\rm Tr}
\left[\log\left({\Delta_{+}+\mu}\right)
-\log\left({\Delta_{-}+\mu}\right)\right]
\label{parametric}
\end{equation}
The amputated determinant appearing in the ratio $R$ is properly
defined as, 
\begin{equation}  
{\rm det'}(\Delta_{-})= \lim_{\mu\rightarrow 0}\left[ \frac{{\rm det}
(\Delta_{-}+\mu)}{\mu^{2k}}\right] 
\label{det'}
\end{equation}
The power of $\mu$ appearing in the denominator reflects the number of
zero eigenvalues of $\Delta_{-}$ calculated above. Using the relation 
${\rm Tr}\log(\hat{O})=\log{\rm det}(\hat{O})$ in (\ref{parametric}), 
we obtain a closed formula for $R$:
\begin{equation}
R=\lim_{\mu\rightarrow 0} \left[\mu^{2k}\exp\left(
\int_{\mu}^{\infty}\,\frac{d\mu'}{\mu'}\, {\cal I}(\mu')\right)\right]^{
\frac{1}{2}}
\label{dresult}
\end{equation}
Evaluating this formula on (\ref{Weinberg}) we obtain the result, 
$R=(2M_{W})^{2k}$. For $k=1$, we can combine the various factors 
(\ref{bmeasure}), (\ref{fmeasure}) and (\ref{cancel}) to obtain the
final result for the one-instanton measure,  
\begin{eqnarray}
\int\,d\mu^{(k=1)} & = & \int\,d\mu_{B}\int\,d\mu_{F}\,R\,\exp(-S_{\rm cl}+i\sigma)
\nonumber \\
& = & \frac{M_{W}}{2\pi} \int\,d^{3}X\,d^2\xi d^2\xi'
\exp(-S_{\rm cl}+i\sigma)
\label{measure2}
\end{eqnarray}
Here we have performed the integration over $\theta$,  anticipating 
the fact that the integrands we are 
ultimately interested in will not depend on this variable. The term
$i\sigma$ is the contribution of the surface term (\ref{surface})
which we will discuss further below. 

\subsection{Instanton effects in the low energy theory}
\paragraph{}   
Because of their long-range fields, instantons and anti-instantons
have a dramatic effect on the low-energy dynamics of three-dimensional
gauge theories. As they can be thought of as magnetic charges in
$(3+1)$-dimensions, 3D instantons and anti-instantons experience a
long-range Coulomb interaction. In the presence of a massless adjoint
scalar, the repulsive force between like magnetic charges is
cancelled. This cancellation is reflected in the existence of static
multi-monopole solutions in the BPS limit. However, in the case of an 
instanton and anti-instanton, there is always an attractive
force. In a purely bosonic gauge theory, Polyakov \cite{pol} 
showed that a dilute 
gas of these objects leads to the confinement of electric charge. The
long-range effects of instantons and anti-instantons can be captured
by including terms in the low-energy effective Lagrangian 
of the characteristic form, 
\begin{eqnarray}
{\cal L}_{I} & \sim & \exp \left(- \frac{8\pi^{2}|\phi|}{e^{2}} \pm 
i\sigma \right) 
\label{si}
\end{eqnarray}
where $|\phi|^{2}=\phi^{2}_{1}+\phi^{2}_{2}+\phi^{2}_{3}$. 
This is simply the instanton action of the previous section, with the
VEVs ${\rm v}_{i}$ and $\sigma$ being promoted to dynamical fields.  
\paragraph{}
In the presence of massless fermions, the instanton contribution to the
low-energy action necessarily couples to $n$ fermion fields, where $n$ is
the number of zero modes of the Dirac operator in the monopole
background. In the three-dimensional $N=2$ theory considered in 
\cite{ahw}, instantons induce a mass term for the 
fermions in the effective action. In fact this corresponds to 
an instanton-induced superpotential  
which lifts the Coulomb branch. 
In the present case, the effective
action will contain an induced four-fermion vertex due to the
contribution of a single instanton \cite{SW3}. 
This vertex has a simple form when
written in terms of the (dimensionally-reduced) Weyl fermions of
$(\ref{lclf})$, 
\begin{eqnarray}
S_{I} & = & \kappa \int \, d^{3}x \, \bar{\lambda}^{2}\bar{\psi}^{2}\,
\exp\left(- \frac{8\pi^{2}|\phi|}{e^{2}} + i\sigma \right) 
\label{sif}
\end{eqnarray}
where the coefficient $\kappa$ will be calculated explicitly below. 
In this case, $N=4$ SUSY does not allow the generation of a 
superpotential and, as we will review below, the four-fermion vertex is 
instead the supersymmetric completion of an 
instanton correction to the metric on the moduli space.   
\paragraph{}
Just like the vertex induced by (four-dimensional) instantons in the
four-dimensional $N=2$ theory, self-duality means that the above 
vertex contains only Weyl fermions of a single chirality. 
Equivalently, the vertex contains 
only the holomorphic components of the 3D Majorana fermions 
in the complex basis 
of (\ref{lclf2}). In the 4D case, the chiral form of the
vertex signals the presence of an anomaly in the $U(1)_{{\cal R}}$
symmetry. By virtue of our dimensional reduction and choice of vacuum, 
the $U(1)_{{\cal
R}}$ symmetry of the 4D theory corresponds to $U(1)_{N}$
defined in Section 1, the unbroken subgroup of $SU(2)_{N}$ global 
symmetry in the 3D theory. More precisely the corresponding charges
are related as $Q_{N}=Q_{{\cal R}}/2$. Hence each of the right-handed 
4D Weyl
fermions carries $U(1)_{N}$ charge $-1/2$, which suggests that the
induced vertex (\ref{sif}) violates the conservation of $U(1)_{N}$ by
$-2$ units. However, as explained in \cite{SW3}, the 
symmetry is non-anomalous in 3D due to the presence of the surface
term $i\sigma$ in the instanton exponent. Assigning $\sigma$ the
$U(1)_{N}$ transformation, 
\begin{eqnarray}
\sigma & \rightarrow & \sigma + 2\alpha
\label{trans}
\end{eqnarray} 
under the action of $\exp(i\alpha Q_{N})$, the symmetry of the
effective action is restored. However, because the VEV of $\sigma$
transforms non-trivially, the symmetry is now spontaneously broken and
hence $SU(2)_{N}$ has no unbroken subgroup. An equivalent statement is
that the orbit of $SU(2)_{N}$ on the classical moduli space is
three-dimensional, a fact that will play an important role in the
considerations of the next Section. 
\paragraph{}
In the following we will be interested in the instanton corrections to
the leading terms in the derivative expansion of the low-energy
effective action. As in the four-dimensional theory, this restricts
our attention to terms with at most two derivatives or four fermions. 
However, an important difference with that case 
is that the 3D instantons are exact solutions of the
equations of motion and there is no mechanism for the VEV to lift
fermion zero modes in a way that preserves the $U(1)_{N}$ symmetry
\footnote{A more general analysis of this issue will be presented
elsewhere.}. It follows that the only sectors of the theory which can
contribute to the leading terms in the low-energy effective action are
those with instanton number (magnetic charge) $-1$, $0$ or
$+1$. Another important difference with the 4D case is that,
because there is no holomorphic prepotential in 3D, there can also
be contributions to the effective action from configurations 
containing arbitrary numbers of instanton/anti-instanton pairs. 
\paragraph{}
Finally we will compute the exact coefficient of the four-anti-fermion
vertex (\ref{sif}) by examining the large distance behaviour of the
correlator 
\begin{eqnarray} 
G^{(4)}(x_{1},x_{2},x_{3},x_{4}) & =& \langle \lambda_{\alpha}(x_{1})   
\lambda_{\beta}(x_{2})\psi_{\gamma}(x_{3})\psi_{\delta}(x_{4}) 
\rangle 
\label{correlator}
\end{eqnarray}
where the Weyl fermion fields are replaced by their zero-mode values in
the one-instanton background. 
The explicit formulae for the zero-modes of the low-energy fermions 
are straightforward to  
extract, via (\ref{fmodes}) from the long-range behaviour of the
magnetic monopole fields. Using the formulae of Appendix C for the 
large-distance (LD) limit of the fermion zero modes we find, 
\begin{eqnarray}
\lambda^{\rm LD}_{\alpha} & =  & 8\pi 
\left(\S_{\rm F}(x-X)\right)_{\alpha}^{\ \beta}\xi_{\beta} \nonumber \\
\psi^{\rm LD}_{\alpha} & =  & 8\pi 
\left(\S_{\rm F}(x-X)\right)_{\alpha}^{\ \beta}\xi'_{\beta}
\label{ld}
\end{eqnarray} 
This asymptotic form is valid for 
$|x-X| \gg  M_{W}^{-1}$ where $X$ is the instanton position and
$\S_{\rm F}(x)=\gamma_\mu x_\mu/(4\pi |x|^{2})$ is the three-dimensional Weyl fermion
propagator. The leading semiclassical contribution to the  correlator 
(\ref{correlator}) is given by,  
\begin{eqnarray} 
G^{(4)}(x_{1},x_{2},x_{3},x_{4}) 
& =& \int\, d\mu^{(k=1)} \lambda^{\rm LD}_{\alpha}(x_{1})   
\lambda^{\rm LD}_{\beta}(x_{2})\psi^{\rm LD}_{\gamma}(x_{3})
\psi^{\rm LD}_{\delta}(x_{4})
\label{corr2}
\end{eqnarray}
where $d\mu^{(k=1)}$ is the one-instanton measure (\ref{measure2}). 
Performing the $\xi$ and $\xi'$ integrations we obtain, 
\begin{eqnarray}
G^{(4)}(x_{1},x_{2},x_{3},x_{4}) & =& 
2^{9}\pi^{3}M_{W}\exp(-S_{\rm cl}+i\sigma) \int\,d^{3}X\, 
\epsilon^{\alpha'\beta'}\S_{\rm F}(x_{1}-X)_{\alpha\alpha'}
 \nonumber \\ 
&  \times & \S_{\rm F}(x_{2}-X)_{\beta\beta'}
\epsilon^{\gamma'\delta'}\S_{\rm F}(x_{3}-X)_{\gamma\gamma'}
\S_{\rm F}(x_{1}-X)_{\delta\delta'} 
\label{corr3}
\end{eqnarray}
This result is equivalent to the contribution of the vertex
(\ref{sif}) added to the classical low-energy action
(\ref{lclf}). Our calculation shows that the coefficient $\kappa$
takes the value, 
\begin{equation}
\kappa = 2^{7}\pi^{3}M_{W}\left(\frac{2\pi}{e^{2}}\right)^{4}
\label{final}
\end{equation}
where the four powers of $(2\pi/e^{2})$ reflect our choice of normalization 
for the fermion kinetic terms in (\ref{lclf}).

\section{The Exact Low-Energy Effective Action}
\paragraph{}   
In this Section, following the arguments of \cite{SW3}, we will determine  
the exact low energy effective action of the $N=4$ SUSY gauge theory in
three dimensions. Below, we will write down the most general 
possible ansatz for the terms in the low-energy effective 
action with at most two derivatives or four fermions which is consistent 
with the symmetries of the model. 
As we will review, the combined restrictions of 
$N=4$ supersymmetry and the global $SU(2)_{N}$ symmetry 
lead to a set of non-linear 
ordinary differential equations for the components of the hyper-K\"{a}hler 
metric which in turn 
determines the relevant terms in the effective action. We will 
find a one-parameter family of solutions of these equations 
which agree with the one-loop 
perturbative calculation of Section 2. Our main result is 
that the one-instanton 
contribution to the four-fermion vertex, 
calculated from first principles in Section 3, uniquely selects the 
metric of the Atiyah-Hitchin manifold from this family of solutions.       
\paragraph{}
We begin by discussing the general case  
of a low-energy theory with scalar fields 
$\{X_{i}\}$ and Majorana\footnote{
The 3D Majorana condition is $\bar{\Omega} = i\Omega$, see Appendix A.} 
fermion superpartners 
$\{\Omega^{\alpha}_{i}\}$ where $i=1,\ldots,d$. As usual the scalars
define coordinates on the quantum moduli space, ${\cal M}$, which is a
manifold of real dimension $d$. The low-energy effective
action has the form of a
three-dimensional supersymmetric non-linear $\sigma$-model with ${\cal
M}$ as the target manifold, 
\begin{equation}
S_{\rm eff}=K \int\,d^{3}x \ \left\{\hf g_{ij}(X) \ 
\left[\partial_{m}X^{i}\partial_{m}X^{j}
-\Omega^{i}\ssl{D}\Omega^{j}\right] -
\twe R_{ijkl}(\Omega^{i}\cdot\Omega^{k})(\Omega^{j}\cdot\Omega^{l})
\right\}
\label{seff1}
\end{equation}
where $K$ is an overall constant included for later convenience. 
The kinetic terms in the action  
define a metric $g_{ij}$ on the moduli space ${\cal M}$, and $\ssl{D}$
and $R_{ijkl}$ denote the corresponding covariant Dirac operator and
Riemann tensor respectively. 
\paragraph{}
Following Alvarez-Gaume and Freedman \cite{ag}, 
the supersymmetries admitted by the above action are written in the
form, 
\begin{equation}
\delta X_{i}=\epsilon^{[1]}\cdot \Omega_{i}+\sum_{q=2}^{N} J^{[q]j}_{i}(X)  
\epsilon^{[q]}\cdot \Omega_{j}
\label{susy}
\end{equation}
The action (\ref{seff1}) is automatically invariant under the 
$N=1$ supersymmetry parametrized by $\epsilon^{[1]}$. However, $N>1$
supersymmetry requires the existence of $N-1$ linearly independent
tensors $J_{i}^{[q]j}$ which commute with the infinitesimal
generators of the holonomy group $H$ of the target. 
It is also necessary that these tensors form an $su(2)$ algebra. 
In general the 
holonomy group of a $d$-dimensional Riemannian manifold is a subgroup 
of $SO(d)$. For $N=4$, the existence of three such 
tensors which commute with the
holonomy generators imply that $d$ must divisible by $4$ and that 
$H$ is restricted to be a subgroup of $Sp(d/4)$. 
Such manifold of symplectic holonomy is by
definition hyper-K\"{a}hler. The tensors $J_{i}^{[q] j}$ ($q=2,3,4$)
define three inequivalent complex structures on ${\cal M}$. 
An equivalent statement of the hyper-K\"{a}hler condition is to require that 
these complex structures be covariantly constant with respect to the
metric $g_{ij}$.   
\paragraph{}
In the case $d=4$, the holonomy can be
chosen to lie in the $Sp(1)\simeq SU(2)$ subgroup of $SO(4)$ generated
by the self-dual tensors $\eta^{a}_{ij}$ $a=1,2,3$ (defined in
Appendix A). In fact the holonomy 
generators are components of the Riemann tensor $R_{ijkl}$ and a
sufficient condition for  a hyper-K\"{a}hler four-manifold  is 
for this tensor to be self-dual\footnote{The more 
conventional choice of an anti-self-dual 
Riemann tensor and self-dual complex structures is related to this one 
by a simple redefinition of the fields and the parameters 
$\epsilon^{[q]}$.}. 
Correspondingly the complex structures $J_{i}^{[q]j}(X)$ ($q=2,3,4$) 
can be taken as linear 
combinations of the anti-self-dual 
$SO(4)$ generators $\bar{\eta}^{a}_{ij}$. As indicated, the relevant 
linear combinations appearing in (\ref{susy}) 
will vary as one moves from one point on the manifold to another. 
Another restriction on the moduli space ${\cal M}$ comes from the
action of the global $SU(2)_{N}$ symmetry. As we have
seen above, there is no anomaly in this symmetry either in 
perturbation theory or from non-perturbative effects, hence we expect
that the exact quantum moduli space has an
$SU(2)_{N}$ isometry. Further, because of the non-trivial
transformation of the dual photon described above, we know that
$SU(2)_{N}$ will generically have three-dimensional orbits on ${\cal
M}$. It can also be checked explicitly that the three complex
structures on ${\cal M}$ introduced above transform as a ${\bf 3}$ of
$SU(2)_{N}$.  
\paragraph{}
Remarkably, the problem of classifying all 
the hyper-K\"{a}hler manifolds of dimension four with the required
isometry has been solved in an entirely different physical context. 
As discussed in Section 1, these are exactly the properties of the
reduced or centered moduli space of two BPS monopoles of gauge group
$SU(2)$. Atiyah and Hitchin \cite{AH} considered all manifolds with these
properties and showed that there is only one manifold which has no 
singularities: the AH manifold. In the original context, the absence
of singularities was required because of known properties of multi-monopole 
solutions. In particular, the metric on the moduli space of an
arbitrary number of BPS monopoles was known to be  complete. 
This means that every curve of finite length on the manifold has a
limit point (see Chapter 3 of \cite{AH} and references therein).  
In the present context, the absence of singularities on the quantum
moduli space ${\cal M}$ can be taken as an assumption about the
strong-coupling behaviour of the 3D SUSY gauge theory. This
assumption leads directly to Seiberg and Witten's proposal that the
quantum moduli space has the same hyper-K\"{a}hler metric 
as the  the AH manifold. 
In the following we will show that the
one-instanton contribution, Eqs (\ref{final}) and (\ref{sif}), 
calculated in the previous section, together with 
the results of one-loop perturbation theory (\ref{1loop}), 
provides a direct proof of the SW proposal without assuming the 
absence of strong-coupling singularities. 
\paragraph{}
Following Gibbons and Manton \cite{gm}, we parametrize the orbits of 
$SO(3)_{N}$ (whose double-cover is $SU(2)_{N}$) 
by Euler angles $\theta$, $\phi$ and $\psi$ with ranges,  
$0\leq \theta \leq \pi$, $0\leq \phi <2\pi$ and  $0\leq \psi <2\pi$
and introduce the standard left-invariant one-forms, 
\begin{eqnarray}
\sigma_{1} & =&-\sin\psi \ d\theta+\cos\psi\sin\theta \ d\phi \nonumber \\     
\sigma_{2} &= &\cos\psi \ d\theta+\sin\psi\sin\theta \ d\phi \nonumber \\   
\sigma_{3} &= &d\psi+\cos\theta \ d\phi       
\label{1forms}
\end{eqnarray}             
The remaining dimension of the moduli space ${\cal M}$, transverse to
the orbits of $SO(3)_{N}$, is labelled by a parameter $r$ and the
most general possible metric with the required isometry takes the form 
\cite{gp},
\begin{equation}
g_{ij}dX_{i}dX_{j}=f^{2}(r)dr^2+a^{2}(r)\sigma_{1}^{2}+
b^{2}(r)\sigma_{2}^{2}+ c^{2}(r)\sigma_{3}^{2}    
\label{metric}
\end{equation}
The function $f(r)$ depends on the definition of the radial parameter
$r$. We will also define the
corresponding Cartesian coordinates, 
\begin{eqnarray}
X & =& r\sin\theta\cos\phi \nonumber \\
Y & =& r\sin\theta\sin\phi \nonumber \\
Z & =& r\cos\theta  
\label{cart}
\end{eqnarray}
\paragraph{}
We start by identifying the parameters introduced above in terms of the 
low-energy fields of Section 2.2 in a way which is consistent with their 
transformation properties under $SO(3)_{N}$. 
The triplet $(X,Y,Z)$ 
transforms as a vector of $SO(3)_{N}$ and will
be identified, up to a rescaling,   
with the triplet of scalar fields appearing in
the Abelian classical action (\ref{lcl}):  
$(X,Y,Z)=(S_{\rm cl}/M_{W})(\phi_{1},\phi_{2},\phi_{3})$. 
This means that we are defining the parameter 
$r$ to be equal to $S_{\rm cl}$.  
The remaining Euler angle, $\psi$, by definition 
transforms as $\psi \rightarrow \psi+
\alpha$ under a rotation through an angle $\alpha$ about the axis
defined by the vector $(X,Y,Z)$. After 
comparing this transformation property with 
the $U(1)_{N}$ transformation (\ref{trans}) of the dual photon
$\sigma$, we set $\psi=\sigma/2$. In the following we will refer to 
$(X,Y,Z,\sigma)$ as standard coordinates.   
\paragraph{}
The weak-coupling behaviour of the metric $g_{ij}$ 
can be deduced by comparing 
the general low-energy effective action 
(\ref{seff1}) with its classical counterpart 
(\ref{lcl}), together with the 
one-loop correction (\ref{1loop}).  
Choosing the constant $K$ in (\ref{seff1}) to be equal to
$2e^{2}/(\pi(8\pi)^{2})$, the classical metric is just the flat one, 
$\delta_{ij}$, in the standard coordinates introduced above. 
Including the one-loop renormalization of the coupling (\ref{1loop}), 
the metric functions, $a^{2}$, $b^{2}$ and $c^{2}$ are given by,
\begin{eqnarray}
  a^{2} = b^{2} & \simeq &  S_{\rm cl}(S_{\rm cl}-2)   \nonumber \\
          c^{2} & \simeq &  4+8/S_{\rm cl}
\label{abc}      
\end{eqnarray}
Corrections to the RHS of the above formulae come from two-loops
and higher and are down by powers of $1/S_{\rm cl}$.  
In fact, the equality $a^{2}=b^{2}$ persists to all orders in perturbation 
theory, reflecting the fact that $U(1)_{N}$ is only 
broken (spontaneously) by non-perturbative effects. 
To this order, the function $f^{2}$ is also determined to be equal to 
$1-2/S_{\rm cl}+O(1/S_{\rm cl}^{2})$.  
\paragraph{}
The Majorana fermions ${\Omega}^{i}_{\alpha}$ 
appearing in (\ref{seff1}), will be real linear combinations of the 
fermions $\chi^{A}_{\alpha}$ of the low-energy action (\ref{lclf}).  
In general, this linear relation will 
have a non-trivial dependence on the bosonic 
coordinates. For our purpose it will be sufficient to determine this 
relation at leading order in the weak-coupling limit, 
$r=S_{\rm cl}\rightarrow\infty$. In the standard coordinates we set,    
\begin{eqnarray}
\Omega^{i}_{\alpha} &\simeq & M^{iA}(\theta,\phi,\sigma)\chi_{A\alpha}      
\label{linear}
\end{eqnarray}
up to corrections of order $1/S_{\rm cl}$. It is convenient to 
complexify the $SO(4)_{\cal R}$ 
index as in Section 2.2 and consider instead coefficients $M^{ia}$ and 
$M^{i\bar{a}}=(M^{ia})^{*}$. In order to reproduce the fermion kinetic term 
in (\ref{lclf2}) these coefficients must obey the relations,  
\begin{eqnarray}
\delta_{ij} M^{ia}M^{j\bar{b}} & = &\delta^{a\bar{b}}
\left(\frac{S_{\rm cl}}{M_{W}}\right)^{2} \nonumber \\
\delta_{ij}M^{ia}M^{jb} & = & \delta_{ij}M^{i\bar{a}}M^{j\bar{b}}\,\, =\,\, 0
\label{const}
\end{eqnarray} 
The first condition fixes the overall normalization of the fermions in 
(\ref{seff1}) while 
the second is required for their kinetic term to be $U(1)_{N}$ invariant. 
The remaining freedom present in this  
identification is related to the ${\cal R}$-symmetry of the 
$N=4$ SUSY algebra. 
\paragraph{}
The hyper-K\"{a}hler condition 
can be formulated as a set of non-linear ordinary
differential equations for the functions $a$, $b$, $c$ and $f$:  
\begin{equation}
\frac{2bc}{f}\frac{da}{dr}=(b-c)^{2}-a^{2}
\label{ode}
\end{equation} 
together with the two equations obtained by cyclic permutation of $a$,
$b$ and $c$. The solutions of these equations are analysed in detail
in Chapter 9 of \cite{AH} and we will adapt the analysis given there
to our current purposes. The equations completely determine the
behaviour of $a$, $b$ and $c$ as functions of $r$ only once one makes
a specific choice for the function $f$, for example the choice
$f=-b/r$ made by Gibbons and Manton \cite{gm}. 
If one makes such a choice in the present context, then the 
corresponding relation between $r$ 
and the weak-coupling parameter $S_{\rm cl}$ 
will receive quantum corrections 
which cannot be determined. In fact we have chosen instead to define $r$ 
to be equal to $S_{\rm cl}$ and, correspondingly, 
this implies some choice for $f$ which can only be determined order by order 
in perturbation theory. This reflects an
important feature of the exact results for the low-energy structure
of SUSY gauge theories which is familiar from
four-dimensions. In these theories the constraints of supersymmetry and (in
the 4D case) duality allow one to specify the exact 
quantum moduli space as a Riemannian manifold. 
However the Lagrangian fields and
couplings of the conventional weak-coupling description 
define a particular coordinate system on the manifold 
and sometimes the 
relation of these parameters to
the parameters of the exact low-energy effective Lagrangian can not be
determined explicitly. As discussed in \cite{dkm2}, just such an
ambiguity arises in the relation between the tree-level coupling constant 
and the exact low-energy coupling in the finite four-dimensional 
$N=2$ theory with four hypermultiplets.  
\paragraph{}
In the light of the above discussion we 
will eliminate both $f$ and $r$ from the equations, and focus on the 
information about the metric which is independent of the choice of
parametrization. Following \cite{AH} 
we obtain a single differential equation for $x=b/a$ and
$y=c/a$,
\begin{equation}
\frac{dy}{dx}=\frac{y(1-y)(1+y-x)}{x(1-x)(1+x-y)}
\label{xyeqn}
\end{equation}
The solutions of this equation are described by curves or trajectories 
in the $(x,y)$-plane.  
In the weak coupling limit $S_{\rm cl}\rightarrow\infty$ we have  
$a^{2}=b^{2}\rightarrow \infty$ and $c^{2}\rightarrow 4$. Hence we must find 
all the solutions of (\ref{xyeqn}) which pass
through the point $Q$ with coordinates 
$x=1$, $y=0$ (see Diagram 7 on p74 of \cite{AH}). 
The relevant features of these solutions are as follows:  
\paragraph{}
{\bf 1:} A particular solution which passes through $Q$ is the
trajectory $x=1$. Returning to the full equations (\ref{ode}),
we find that this corresponds to the solution: 
\begin{equation}
a=b=\frac{c}{(1-c^{2}/4)} 
\label{ptsoln}
\end{equation}
Eliminating $S_{\rm cl}$ in (\ref{abc}), we find this relation is obeyed
up to one-loop in perturbation theory as long as we choose square
roots so that $ac=bc<0$. This solution describes
the singular Taub-NUT geometry and, as discussed in \cite{SW3}, this is the
exact solution of the low-energy theory up to non-perturbative
corrections. In other words the resulting effective action (\ref{seff1}) 
includes the sum of all corrections from all orders in 
perturbation theory. However, as we have commented above, the function $f$ 
is not determined by the solution, 
so the all-orders effective action cannot be written explicitly in terms of 
the weak coupling parameter $S_{\rm cl}$.       
\paragraph{}
{\bf 2:} There is precisely a one-parameter family of solutions
passing through $Q$, each of which is exponentially close to the line
$x=1$ ($y<0$) near $Q$. Linearizing around $x=1$, $y=0$ we may
integrate (\ref{xyeqn}) to obtain the leading asymptotic behaviour,  
\begin{eqnarray}
a-b & \simeq & B \frac{a^{2}}{c}\exp\left(\frac{2a}{c}\right)
\label{asymptotic}
\end{eqnarray}
where $B$ is a constant of integration. Corrections to the RHS are
down by powers of $y=c/a$ or by powers of $\exp(2/y)$. 
\paragraph{}
{\bf 3:} Using numerical methods to examine 
the behaviour of these solutions away from the point $Q$,
one finds that there is a unique critical trajectory which originates at the
point $P'$ with coordinates $(0,-1)$. All other trajectories originate
either at the origin $(0,0)$ or at negative infinity $(0,-\infty)$. 
Atiyah and Hitchin show that only the critical trajectory corresponds
to a complete manifold. The critical solution can be constructed
explicitly in terms of elliptic functions, its asymptotic form near
$Q$ is given in Gibbons and Manton \cite{gm} 
(see equation (3.14) of this reference) and agrees with (\ref{asymptotic})
with a specific value for the integration constant
$B=B_{cr}=16\exp(-2)$. All trajectories with $B\neq B_{cr}$ correspond
to singular geometries. 
\paragraph{}
Using the identifications (\ref{abc}), 
the asymptotic behaviour (\ref{asymptotic}) becomes, 
\begin{eqnarray}
a-b & \simeq & -8q S_{\rm cl}^{2}\exp\left(-S_{\rm cl}\right) 
\label{prediction} 
\end{eqnarray}
where $q=B/B_{cr}$. Hence the leading deviation from the 
perturbative relation $a=b$ 
comes with exactly the exponential suppression characteristic of a 
one-instanton effect. When substituted in the metric (\ref{metric}) and 
the effective action (\ref{seff1}), this term yields a contribution 
to the boson kinetic terms which also comes with the phase factor  
$\exp(\pm i\sigma)$  expected for the (anti-)instanton term. 
Further, each member of the family of solutions parametrized by the constant 
$q$ yields a different prediction for this one-instanton effect.    
In principle, it 
is straightforward to check the coefficient of this term against the 
results of a semiclassical calculation of the scalar propagator. 
This would involve calculating the Grassmann bilinear contributions to the 
scalar field which come from the fermion zero modes in the monopole 
background. In the 
following, we will choose instead to extract a prediction for the 
four-fermion vertex in (\ref{seff1}) 
and compare this directly with the result (\ref{final}) of Section 3. 
\paragraph{}
The effective action (\ref{seff1}) contains a four-fermion vertex 
proportional to the Riemann tensor. To make contact with the results of 
Section 3 where the vacuum is chosen to lie in the $\phi_{3}$ direction in 
orbit of $SU(2)_{N}$, we evaluate 
the Riemann tensor corresponding to the metric (\ref{metric}) 
at the point on the manifold 
with standard coordinates $(0,0,r=S_{\rm cl},\sigma)$. 
This calculation is presented in Appendix D (Many of the necessary results 
have been given previously by Gauntlett and Harvey 
in Appendix B of \cite{gh}).  In particular, we 
evaluate the leading-order contribution to the $\sigma$-dependent terms in 
the Riemann tensor in the weak-coupling limit: $S_{\rm cl}\rightarrow\infty$. 
The result is best expressed in the complex basis with coordinates, 
\begin{eqnarray}
z_{1}=\frac{1}{\sqrt{2}}(X-iY) & \qquad{} \qquad{} \qquad{} & 
z_{2}=\frac{1}{\sqrt{2}}(Z- i\sigma)
\label{cbasis}
\end{eqnarray}
In this basis, the relevant contribution to the Riemann tensor is 
pure holomorphic and is given by, 
\begin{equation}
R_{1212}=8qS_{\rm cl}\exp\left(-S_{\rm cl}+i\sigma\right)       
\label{rtensor}
\end{equation}
where the other pure holomorphic components are related to this one by the 
usual symmetries of the Riemann tensor: 
$R_{abcd}=-R_{bacd}=R_{cdab}=-R_{abdc}$. The anti-instanton contribution is 
pure anti-holomorphic and all components of mixed holomorphy are 
independent of $\sigma$. 
\paragraph{}
The above result for the Riemann tensor means that the 
corresponding four-fermion vertex in the non-linear $\sigma$-model 
(\ref{seff1}), has exactly the chiral form expected 
from the discussion of Section 3.2. In particular, invariance of the 
vertex under $U(1)_{N}$ transformations 
implies that the holomorphic (anti-holomorphic) components $\Omega^{a}$ 
($\Omega^{\bar{a}}$) of the target space fermions  
have $U(1)_{N}$ charge $-1/2$ ($+1/2$). 
Hence we complete our identification of the fermions by demanding that the 
transformation (\ref{linear}) maps fermions of positive (negative) 
$U(1)_{N}$ charge to fermions of positive (negative) $U(1)_{N}$ charge. 
At the chosen point 
$(0,0,r=S_{\rm cl},\sigma)$, we write the relation between fermions,  
($\Omega^{a},\Omega^{\bar{a}}$) in the complex basis (\ref{cbasis}) 
and ($\chi^{b},\chi^{\bar{b}}$) in the complex basis of (\ref{lclf2}) as, 
\begin{eqnarray} 
\Omega_{a}^{\alpha} & \simeq & M_{ab}\chi_{b}^{\alpha} \nonumber \\
\Omega_{\bar{a}}^{\alpha} & \simeq & M_{\bar{a}\bar{b}}\chi_{\bar{b}}^{\alpha}
\label{linear2}
\end{eqnarray}
where $M_{\bar{a}\bar{b}}=(M_{ab})^{*}$ and, as in (\ref{linear}), 
corrections to the RHS are down by inverse powers of $S_{\rm cl}$.  
The second condition in (\ref{const}) is automatically satisfied. 
Regarding $M_{ab}$ as a $2\times 2$ matrix, the first condition in 
(\ref{const}) is satisfied by choosing $M=(S_{\rm cl}/M_{W})\hat{M}$, where 
$\hat{M}\in SU(2)$ is a residual degree of freedom associated with the 
unbroken $SU(2)_{\cal R}$ symmetry\footnote{Strictly speaking $M$ is only 
restricted to lie in $U(2)$. However the additional phase can be reabsorbed 
by changing the identification $\psi=\sigma/2$ made above 
by an additive constant.}.    
\paragraph{}
Finally we calculate the four-fermion vertex which follows from the 
instanton contribution to the Riemann tensor (\ref{rtensor}).  
After taking into account the 
symmetries of the Riemann tensor described above 
and performing a Fierz rearrangement the resulting vertex becomes, 
\begin{equation}
{\cal L}_{\rm{4F}}=\frac{1}{4}K\,R_{1212}
\left(\Omega^{1}\cdot\Omega^{1}\right)
\left(\Omega^{2}\cdot\Omega^{2}\right)
\label{4f1}
\end{equation}
We rewrite the vertex in terms of Weyl fermions using, 
\begin{eqnarray}
\left(\Omega^{1}\cdot\Omega^{1}\right)
\left(\Omega^{2}\cdot\Omega^{2}\right) & = & \left({\rm det}(M)\right)^{2}
\left(\chi^{1}\cdot\chi^{1}\right)
\left(\chi^{2}\cdot\chi^{2}\right) \nonumber \\
& =& \left(\frac{S_{\rm cl}}{M_{W}}\right)^{4}\bar{\lambda}^{2}\bar{\psi}^{2}
\label{ff}
\end{eqnarray}
Collecting together the various factors, the final result for the 
induced four-fermi vertex is,
\begin{equation}
{\cal L}_{\rm{4F}}=2^{7}\pi^{3}qM_{W}\left(\frac{2\pi}{e^{2}}\right)^{4}\,
\bar{\lambda}^{2}\bar{\psi}^{2}\,\exp\left(-S_{\rm cl}+i\sigma\right) 
\label{final2}
\end{equation}
Comparing this with the calculated value (\ref{final}) for 
the coefficient $\kappa$ in the instanton-induced vertex (\ref{sif}), 
we deduce that $q=1$. This implies that the quantum moduli space of the 
theory is in fact the Atiyah-Hitchin manifold.
\paragraph{}
\centerline{******************}

The authors would like to thank C. Fraser, N. Manton, B. Schroers 
and E. Witten for useful discussions. 
ND, VVK and MPM would like to thank the Isaac 
Newton Institute for hospitality while part of this work was completed.  
ND is supported by a PPARC Advanced Research Fellowship. 
MPM is supported by the Department of Energy. 

\section*{Appendix A: Dimensional Reduction}

In this Appendix we present our conventions and give the details
of dimensional reduction. We work in Minkowski\footnote{With the exception
of this Appendix and the next one, the calculations in the rest of the paper
are performed in Euclidean space} space in 4D and 
3D with the metric signature $(+,-,-,...)$ and $m,n=0,1,2,3$,
$\ \mu,\nu=0,1,2$.
\paragraph{}
We start with the $N=2$ supersymmetric Yang-Mills in 
4D
\begin{eqnarray}
S_{\rm 4D}\ = \ {1\over g^2} \int d^4 x \ 
{\rm Tr}&\Big\{&-\hf\uv_{mn}\uv^{mn}
+i\ulambdabar\Dbarslash\ulambda
+i\ulambda\Dslash\ulambdabar
+\uD^2\\
&&+2\D_m\uAbar\D^m\uA
+i\upsibar\Dbarslash\upsi
+i\upsi\Dslash\upsibar
+2\uFbar\uF\nonumber\\
&&-2\uD\,[\,\uA\,,\uAbar\,]\,+2\sqrtwo 
i\Big(\,[\,\uAbar,\upsi\,]\,\ulambda
+\ulambdabar\,[\,\uA\,,\upsibar\,]\,\Big)\Big\}\ . \nonumber
\label{s4d}
\end{eqnarray}
Here 
 $\uv_m$ is the gauge field, $\uA$ is the complex scalar field, 
Weyl fermions 
$\ulambda$ and $\upsi$ are their superpartners,
 while $\uD$ and $\uF$ are auxiliary fields. 
Also $\Dslash_{\alpha\dot\alpha}\ =\ \D_m\sigma^m_{\alpha\dot\alpha}$, and
$\Dbarslash^{\dot\alpha\alpha}\ =\ \D^m\sigmabar_m^{\dot\alpha\alpha}$,
where $\,\D_m{\uX}\ =\ \partial_m{\uX}-i\,[\,\uv_m,{\uX}\,]$.
Wess and Bagger \cite{WB} spinor summation conventions are used throughout and
sigma-matrices in Minkowski space are,
$\sigma^m_{ \ \alpha\dot\alpha} = ( -1, \tau^a)$, 
$\sigmabar^{m \ \dot\alpha\alpha} =  ( -1, -\tau^a)$.
\paragraph{}
The three-dimensional theory is obtained by making all the fields
independent of one spatial dimension and decoupling this dimension
from the theory, Eq. $(\ref{coupling})$. 
The only subtle point in this program is the dimensional reduction of the
fermions.
In 4D
the 2-component Weyl spinors can be combined into the 4-component Majorana
spinors,
\begin{eqnarray}
\ulambda_{\rm ch}=\pmatrix{\ulambda_\alpha \cr \ulambdabar^{\dot \alpha}} \ ,
\qquad 
&&\ulambdabar_{\rm ch}=(\ulambda^\alpha \ , \ \ulambdabar_{\dot \alpha}) \ ,
\nonumber\\
\upsibar_{\rm ch}=(\upsi^\alpha \ , \ \upsibar_{\dot \alpha})  \ , \qquad 
&&
\upsi_{\rm ch}=\pmatrix{\upsi_\alpha \cr \upsibar^{\dot \alpha}} \ ,
\label{fch}
\end{eqnarray}
in such a way that,
\begin{equation}
i\ulambdabar\Dbarslash\ulambda
+i\ulambda\Dslash\ulambdabar \ = \ 
i\ulambdabar_{\rm ch}\Gamma_{\rm ch}^m \D_m\ulambda_{\rm ch} \ ,
\end{equation}
where $\Gamma_{\rm ch}^m$ is the gamma matrix of the 4D theory
in the standard chiral basis,
\begin{equation}
\Gamma_{\rm ch}^m \ = \ \pmatrix{0 & \sigma^m \cr \sigmabar^m & 0} \ .
\label{gch}
\end{equation}
Thus, $\ulambda_{\rm ch}$ and $\upsi_{\rm ch}$ are the Majorana
fermions in the chiral basis. 
\paragraph{}
For the purposes of dimensional reduction it is more convenient to choose 
a different -- real -- basis for Majorana spinors in 4D
related by a unitary transformation to the chiral basis above,
\begin{eqnarray}
\ulambda_{\rm re}=\pmatrix{\uchi_\alpha \cr \uchitilde_\alpha} \ , \qquad 
&&\ulambdabar_{\rm re}=\  i(\uchitilde^\alpha \ , \ \uchi^\alpha)
 \ ,\nonumber\\
\upsibar_{\rm re}=\ i(\uetatilde^\alpha \ , \ \ueta^\alpha)    \ ,
&&\upsi_{\rm re}=\pmatrix{\ueta_\alpha \cr \uetatilde_\alpha} \ , 
\label{fre}
\end{eqnarray}
where new (real) 2-spinors are simply the `real' and `imaginary'
parts of the Weyl 2-spinors,
\begin{eqnarray}
\uchi_\alpha=\frac{1}{\sqrt 2}(\ulambda_\alpha+\ulambdabar_{\dot \alpha})
\qquad &&\uchitilde_\alpha=-\frac{i}{\sqrt 2}(\ulambda_\alpha-
\ulambdabar_{\dot \alpha})\nonumber\\
\ueta_\alpha=\frac{1}{\sqrt 2}(\upsi_\alpha+\upsibar_{\dot \alpha}) \qquad
&&\uetatilde_\alpha=-\frac{i}{\sqrt 2}(\upsi_\alpha-\upsibar_{\dot \alpha})\ .
\label{ma3}
\end{eqnarray}
The 4D gamma matrices in this basis are
\begin{eqnarray}
\Gamma_{\rm re}^0 = &&\pmatrix{0 & -\tau^2 \cr -\tau^2 & 0} \ , \quad 
\Gamma_{\rm re}^1 = \pmatrix{0 & i\tau^3 \cr i\tau^3 & 0} \ ,
\nonumber\\ 
\Gamma_{\rm re}^2 = &&\pmatrix{i1 & 0 \cr 0 & -i1} \ , \quad 
\Gamma_{\rm re}^3 = \pmatrix{0 & -i\tau^1 \cr -i\tau^1 & 0} \ . 
\label{gre}
\end{eqnarray}
Now the dimensional reduction from 4D to 3D is straightforward,
one has to decouple the second dimension,
$(x_0,x_1,x_2,x_3) \to (x_0,x_1,x_3) \equiv (y_0, y_1, y_2)$.
The four real 2-spinors $\uchi_\alpha$, $\uchitilde_\alpha$
and  $\ueta_\alpha$, $\uetatilde_\alpha$
become Majorana spinors in 3D and the three gamma 
matrices satisfying the Clifford algebra in 3D can be read off from 
$(\ref{gre})$: $\gamma^0=\tau^2, \gamma^1=-i\tau^3, \gamma^2=i\tau^1$.  
Note that the 3D Majorana condition 
$\psi^TC=\psi^\dagger\gamma_0$ is now the condition that the spinor 
is real since the 
charge conjugation matrix is $C=\tau^2$. We 
have defined the Dirac conjugate in 3D as ${\bar \psi}=\psi^\dagger\gamma_0$. 
For a Majorana spinor this means $\uchibar^\alpha=i\uchi^\alpha$.
\paragraph{}
By renaming the gamma matrices in 4D we can always choose the third
and not the second dimension to decouple. This will always be assumed,
see the footnote on page 4.
\paragraph{}
Finally we define bosonic fields $\uphi_{1,2,3}$ and $\uv_\mu$ in 3D as follows,
\begin{equation}
\uv_m = \left\{ \begin{array}{rr}
\uv_\mu \ , &
 m= 0,1,2 \\ \uphi_3 \ , & m=3 
\end{array}\right. \ ,
\qquad
\uA \ = \ {\uphi_1 + i \uphi_2 \over \sqrt{2}}
\ ,
\label{df3}
\end{equation}
and the 4D action $(\ref{s4d})$ becomes 
\begin{eqnarray}
S_{\rm 3D} &=& {2\pi \over e^2} \int d^3 x \ 
{\rm Tr}\big\{\,-\hf\uv_{\mu\nu}\uv^{\mu\nu}+
\D_\mu\uphi_i\D^\mu\uphi_i
-\uchi\Dhat\uchi
-\uchitilde\Dhat\uchitilde
-\ueta\Dhat\ueta
-\uetatilde\Dhat\uetatilde
\nonumber\\
&&+2\uphi_3\big([\,\uchi,\uchitilde\,]+[\,\ueta,\uetatilde\,]\big)
+2\uphi_2\big([\,\ueta,\uchi\,]+[\,\uchitilde,\uetatilde\,]\big)
+2\uphi_1\big([\,\uchi,\uetatilde\,]+[\,\uchitilde,\ueta\,]\big)\ \nonumber\\
&&+\big([\uphi_1,\uphi_2]^2+[\uphi_2,\uphi_3]^2+[\uphi_3,\uphi_1]^2\big)
\big\}
\ .
\label{s3d}
\end{eqnarray}
Here $\Dhat_\alpha^{\ \beta}=\D_\mu\,(\gamma^\mu)_\alpha^{\ \beta}$
with $\D^\mu=\partial^\mu-i[\uv^\mu,\ \,]$ and $\gamma^{0,1,2}$ satisfy
$\{\gamma^\mu,\gamma^\nu\}=2g^{\mu\nu}$.
\paragraph{}
This action can be written in a manifestly $SO(4)_{\cal R}$ invariant form.
First, define $\uchi^A=(\uchi,\uchitilde,\ueta,-\uetatilde)$, where 
$A=1,...,4$ is the $SO(4)_{\cal R}$ index. 
Second, introduce the self-dual and anti-self-dual 
't Hooft $\eta^i$-matrices \cite{th}
\begin{equation}
\eta^{i}_{AB} = \left\{ \begin{array}{rr}
\epsilon_{iAB} &
 A,B= 1,2,3 \\ -\delta_{B i} & A= 4 \\ \delta_{A
i} & B = 4
\end{array}\right.
\qquad
{\bar \eta}^{i}_{AB} = \left\{ \begin{array}{rr}
\epsilon_{iAB} &
 A,B= 1,2,3 \\ \delta_{B i} & A= 4 \\ -\delta_{A
i} & B = 4
\end{array}\right.\ .
\label{eth}
\end{equation}
With this definition, $\eta$ is self-dual and $\bar{\eta}$
anti-self-dual
with respect to $\epsilon^{1234}=+1$. Moreover, they form two 
sets of commuting 
$su(2)$ algebras, as $[\eta^i,{\bar \eta}^j]=0$.
In this notation the action $(\ref{s3d})$ takes the form, 
\begin{eqnarray}
S_{\rm 3D}\ &=& {2\pi \over e^2} \int d^3 x \ 
{\rm Tr}\big\{\,-\hf\uv_{\mu\nu}\uv^{\mu\nu}+\D_\mu
\uphi_i\D^\mu\uphi_i
-\uchi^A \Dhat \uchi^A\nonumber\\
&&+\sum_{i<j}[\uphi_i,\uphi_j]^2+2\uphi_i{\bar \eta}^i_{
AB}\uchi^A\uchi^B \big\}    \ .
\label{sn4}
\end{eqnarray}
The Lagrangian has a global 
$SO(4)_{\cal R}\simeq SU(2)_{N}\times SU(2)_{\cal R}$ 
symmetry. The $SU(2)_{R}$ leaves the three scalar fields invariant, 
and acts on the fermions. In terms of the four-dimensional Weyl fermions 
$(\ulambda\,\,\upsi)$ forms a doublet 
under $SU(2)_R$. Rewriting this in terms of the Majorana's in 3D, one finds
\begin{equation}
\uchi^{A}\mapsto\exp\left(\hf\alpha^{k}\eta^{k}_{
AB}\right)
\uchi^{B} \ .
\end{equation}
The $SU(2)_{N}$ group is the remnant of a 3-dimensional rotation
group in the $N=1$, $D=6$ theory. 
The scalar fields transform as
\begin{equation}
\uphi^{i}\mapsto\exp\left(\beta^{k}R^{k}\right)^{i}{}_{j}\,\uphi^{j}
\ ,
\end{equation}
where $(R^{k})^i{}_j=\epsilon^{kij}, \epsilon^{123}=1$ are the standard rotation 
group generators.
On the fermions, $SU(2)_N$ acts as
\begin{equation}
\uchi^{A}\mapsto\exp\left(\hf\beta^{k}{\bar \eta}^{k}_{AB}\right)
\uchi^{B} 
\end{equation}
These transformations leave the microscopic theory invariant, as one can
check
explicitly. 
\paragraph{}
The supersymmetry transformation rules for the scalar fields are given by
\begin{equation}\label{susy1}
\delta\uphi^{i}=
\bar{\epsilon}^{A\alpha}{\bar \eta}^{i}_{AB}\uchi^{B}_{\alpha}\ 
\end{equation}
They can be obtained from a dimensional reduction of the SUSY rules in 4D. 
One can check that $SU(2)_N$ and $SU(2)_{\cal R}$ are invariances of this 
transformation. Finally we can relate the Majorana fermions $\uchi^{A}$ to 
the (dimensionally-reduced) Weyl fermions $\ulambda$ and $\upsi$ by 
complexifying in the $SO(4)_{\cal R}$ index. We define a complex basis;     
\begin{eqnarray}
\uchi^{1}_{\alpha} = \frac{1}{\sqrt{2}}(\uchi_{\alpha}-i\tilde{\uchi}_{\alpha})
=\epsilon_{\alpha\dot{\beta}}\bar{\ulambda}^{\dot{\beta}} & ; & 
\uchi^{\bar{1}}_{\alpha} = \frac{1}{\sqrt{2}}(\uchi_{\alpha}+i\tilde{\uchi}
_{\alpha})=\ulambda_{\alpha} \nonumber \\  \uchi^{2}_{\alpha} = 
\frac{1}{\sqrt{2}}(\ueta_{\alpha}-i\tilde{\ueta}_{\alpha})
=\epsilon_{\alpha\dot{\beta}}\bar{\upsi}^{\dot{\beta}} & ; & 
\uchi^{\bar{2}}_{\alpha} = \frac{1}{\sqrt{2}}
(\ueta_{\alpha}+i\tilde{\ueta}_{\alpha})=\upsi_{\alpha} 
\end{eqnarray}
\section*{Appendix B: Wilsonian Effective Action at 1-loop}
\def\bkgd{{\sst\rm bkgd}}
%
%
%
%
\paragraph{}
In this Appendix we extract the 1-loop effective $U(1)$ Wilsonian action
from the microscopic $SU(2)$ Lagrangian. 
\paragraph{}
In order to calculate the Wilsonian effective action, it is customary
to split up all fields (except the ghosts)
into a background part and a fluctuating
part, e.g.,
\begin{equation} 
\uphi_i\ =\ \uphi_{i\,\bkgd}+\delta\uphi_i\ , \quad {\rm etc.}
\label{splitup}
\end{equation}
The background  fields should be thought of as comprising
 large-wavelength modes which will
justify a gradient expansion; in particular the background
scalar field includes the VEV {\bf v} which we can choose
to point in the third direction in both colour and $SU(2)_{N}$ space:
\begin{equation}
\uphi_{i\,\bkgd}\ =\ \sqrt{2}\vhiggs\delta_{i3}\tau^3/2+\delta\uphi_{i\,\bkgd}
\label{vevpoint}
\end{equation}
By convention, under an $SU(2)$ gauge transformation,
 the variation of the total field
is entirely assigned to the fluctuating parts while the background parts
are held fixed, thus:
\begin{eqnarray}
\delta_{\theta(x)}\,\delta v_\mu^a\ &= & \
\epsilon^{abc}\theta^b(x)\big(v_{\mu\,\bkgd}^c+\delta v_\mu^c\big)-,
\partial_\mu\theta^a(x) \nonumber \\
&= & \ 
\epsilon^{abc}\theta^b(x)\delta v_\mu^c-(D_\mu\utheta\,)^a\nonumber \\  
\delta_{\theta(x)}\,\delta \phi_k^a\ &= & \ \epsilon^{abc}\theta^b(x)
\left(\phi^c_{k\,\bkgd}+\delta\phi^c_k\right)\ \nonumber \\
\delta_{\theta(x)}\,v_{\mu\,\bkgd}^a\ &=&  \ 
\delta_{\theta(x)}\,\phi_{k\,\bkgd}^a\ =\ 0\ 
\label{gaugeXn}
\end{eqnarray}
and likewise for the fermions.
Here $D^\mu=\partial^\mu-i[\uv^\mu_\bkgd,\ ]$ is the background covariant
derivative. It is convenient to specialize to the one-parameter
family of ``background $R_\xi$ gauges,'' linear
in the fluctuating fields, defined by the gauge-fixing term
\def\Lgft{{\cal L}_{\sst\rm g.f.t.}}
\begin{equation}
\Lgft\ =\ -{2\pi \over e^2}
{1\over2\xi}\sum_{a=1,2,3}\big(f_\xi^a\big)^2
\label{Lgftdef}
\end{equation}
where
\begin{equation}
f_\xi^a\tau^a/2\ =\ D_\mu\,\delta\uv^\mu+i\xi\sum_{k=1,2,3}
[\uphi_{k\,\bkgd}\,,\,\delta\uphi_k]
\label{fxidef}
\end{equation}
As in the usual $R_\xi$
gauges, for any $\xi,$ $\Lgft$ is constructed to cancel out the troublesome
quadratic cross term $-\sqrt{2}
\vhiggs(\delta\phi_3^1\partial^\mu\delta v_\mu^2-
\delta\phi_3^2\partial^\mu\delta v_\mu^1)$ induced in the
$SU(2)$ Lagrangian by
the VEV (\ref{vevpoint}) (with an integration by parts).
The corresponding action for the triplet of complex ghosts $c^i$ follows
straightforwardly from Eq. (\ref{gaugeXn}):
\def\Lghost{{\cal L}_{\sst\rm ghost}}
\begin{eqnarray}
\Lghost\ =  \ {2\pi \over e^2} \ c^{i\dagger}\,  
{\delta f_\xi^i\over\delta\theta^j}\,c^j \qquad \qquad \qquad\qquad 
\qquad \qquad  \qquad \qquad \qquad \qquad  & &\nonumber  \\ 
= \ {2\pi \over e^2} \ 
\vec c^{\,\dagger}\cdot\big[-D^2\ -\ (D^\mu\circ\delta\vec v_\mu\times
\ \ )\ + \ \xi\vec\phi_{k\,\bkgd}\times\big((
\vec\phi_{k\,\bkgd}+\delta\vec\phi_k)\times\ \big)\big]\vec c  && \nonumber \\
\label{Lghostdef}
\end{eqnarray}
using an obvious 3-vector notation for the adjoint fields, 
e.g., $\vec v_\mu=(v_\mu^1,v_\mu^2,v_\mu^3).$
\paragraph{}
We will calculate (part of) the one-loop effective action for the massless
quanta $\{\phi^3_{1\,\bkgd},\phi^3_{2\,\bkgd},\delta\phi^3_{3\,\bkgd},
v^3_{\mu\,\bkgd},\chi^3_{\bkgd},\chitilde^3_{\bkgd},
\eta^3_{\bkgd},\etatilde^3_{\bkgd}\}$. Here
$\phi^3_{1\,\bkgd}$ and $\phi^3_{2\,\bkgd}$
are the Goldstone bosons associated with the spontaneous breaking
of the $SU(2)_{N}$ symmetry down to $U(1)$ under which they transform
as a doublet, whereas $\delta\phi^3_{3\,\bkgd}$ is the singlet dilaton
(cf Eq (\ref{vevpoint})). Since these scalars have different charges under the unbroken $U(1)$, one  generically expects their effective
couplings to renormalize differently at the loop level, hence:
\begin{eqnarray}
\Leff =\frac{2\pi}{ e^2}\frac{1}{2}
\left(1-\frac{C_1e^2}{ M_W}+{\cal O}(e^4)\right)
\sum_{i=1,2}\left(\partial_\mu\phi_{i\,\bkgd}^3\right)^2
\  \qquad \qquad && \nonumber 
\\ \qquad \qquad \qquad \qquad + \frac{2\pi}{ e^2}\frac{1}{2}\left(1-
\frac{C_3e^2}{M_W}+{\cal O}(e^4)\right)
\big(\partial_\mu\delta\phi_{3\,\bkgd}^3\big)^2
\ +\ \cdots\ 
\label{Leffdef}
\end{eqnarray}
where the one-loop 
numerical constants $C_1$ and $C_3$ are not necessarily equal,
and we omit spinor and gauge fields as well as higher derivative terms.
However, since our choice of gauge fixing, (\ref{fxidef}), 
respects both scale and
$SU(2)_{N}$ invariance,  Eq. (\ref{Leffdef}) must come from an $O(3)$-invariant
expression containing no explicit factors of $M_W$ nor of 
$\delta\phi_{3\,\bkgd}^3$; only the total background fields
$\phi_{i\,\bkgd}^3$ can appear. In particular, explicit factors of $M_W$
are replaced by
\begin{equation}
M_W\ \rightarrow\ \rho\ ,\qquad
\rho=\big[\sum_{i=1,2,3}(\phi_{i\,\bkgd}^3)^2\big]^{1/2}
\label{Mreplace}
\end{equation}
Thus (\ref{Leffdef}) must come from 
\begin{eqnarray}
\Leff^\oneloop\ =\ \frac{2\pi}{e^2}\frac{1}{2}
\left(1-\frac{C_1e^2}{\rho}\right)
\sum_{i=1,2,3}\big(\partial_\mu\phi_{i\,\bkgd}^3\big)^2
\ + \qquad \qquad \qquad && \nonumber \\ \qquad \qquad \qquad 
 \frac{2\pi}{ e^2}\frac{(C_1-C_3)e^2}{2\rho^3}\left(\sum_{i=1,2,3}
\phi_{i\,\bkgd}\partial_\mu\phi_{i\,\bkgd}\right)^2\ +\ \cdots &&
\label{Loneloop}
\end{eqnarray}
in terms of the two possible  $SU(2)$ invariants at the 2-derivative level.
The difference between (\ref{Leffdef}) and (\ref{Loneloop}) 
lies in the 3-point and 
higher-point functions when one expands about the VEV.
\paragraph{}
Below we shall explicitly calculate
\begin{equation}
C_1\ =\ C_3\ =\ \frac{1}{4\pi^2}
\label{Cvalues}
\end{equation}
which constitutes our 1-loop prediction.
\paragraph{}
{\bf  Calculation of the effective action.} 
Extracting the 1-loop Wilsonian effective action is an intricate calculation
in a generic ``background $R_\xi$'' gauge, but for the specific value
$\xi=1$ we can exploit the following observation. Let us extend the
gauge field $\uv_\mu$ to a 6-dimensional vector field incorporating the
three scalars $\uphi_i\,$:
\def\sixd{{\sst\rm6D}}
\begin{equation}
\uv_\mu^\sixd\ =\ \big(\,\uv_0\,,\,\uv_1\,,\,\uv_2\,,\,
\uv_3\equiv\uphi_1\,,\,\uv_4\equiv\uphi_2\,,\,\uv_5\equiv\uphi_3\,\big)
\label{vsixdef}
\end{equation}
With the convention that all relevant field configurations are constant
in the final three spatial directions,
\begin{equation}
\frac{\partial}{\partial x^3}\ =\
\frac{\partial}{\partial x^4}\ =\ \frac{\partial}{\partial x^5}\ \equiv\ 0\ 
\label{constancy}
\end{equation}
then the 3-dimensional $N=4$ $SU(2)$ Lagrangian  is known to
be simply
that of 6-dimensional $N=1$ SYM theory. Furthermore, for the specific
choice $\xi=1,$ the gauge-fixing conditions (\ref{fxidef}) may be rewritten
compactly as
\begin{equation}
f_{\xi=1}^a\tau^a/2\ =\ D^\sixd_\mu\,\delta\uv^{\sixd\,\mu}\ 
\label{fxidefsixd}
\end{equation}
where $D_\mu^\sixd$ is the 6-dimensional background covariant derivative
defined subject to (\ref{constancy}), and the metric signature is
$(+,-,-,-,-,-)$. Likewise the ghost action (\ref{Lghostdef}) simplifies to
\begin{equation}
\Lghost\ = \ {2\pi \over e^2} \ 
 \vec c^{\,\dagger}\cdot\big[-(D^\sixd)^2\ -\ D^{\sixd\,\mu}
\circ\delta\vec v_\mu^{\,\sixd}\times\ \ \big]
\vec c
\label{Lghostsixd}
\end{equation}
With these simplifications 
the problem is now reduced, quite literally, to a textbook exercise
(see Sec.~16.6 of \cite{peskin}, which we follow closely).
At the one-loop level we focus solely on terms quadratic in the fluctuating
fields, dropping for example the second term in (\ref{Lghostsixd}). 
The resulting 
Gaussian functional determinants may be exponentiated in the usual
way, and lead to a contribution
\begin{equation}
\sum_{s=0,\frac{1}{2},1}\eta_s\,{\rm Tr}\log\Delta_s
\label{contrib}
\end{equation}
to the effective action. Here $s$ indexes the spin, and the weight
factor $\eta_s$ takes the values $\eta_1=-\hf$ for the 6D
vector, $\eta_{1/2}=+\hf$ for the single 6D Dirac spinor
formed from the four 3D Majorana fermions\footnote{NB: We take 
$\eta_{1/2}=+1/2$ rather than $\eta_{1/2}=+1$ for the spinor 
because in the definition of $\Delta_{1/2}$ we will square the $\Dbarslash$ 
operator following \cite{peskin}.}, and 
$\eta_0=+1$ for the complex ghosts. $\Delta_s$ is the Gaussian quadratic
form sandwiched between the spin-$s$ fluctuating fields in the adjoint
representation of $SU(2)_{\sst\rm color}.$ As shown in \cite{peskin} it has
the universal form:
\def\Deltaone{\Delta^{\sst(1)}}
\def\Deltatwo{\Delta^{\sst(2)}}
\def\Deltatwotilde{\tilde\Delta^{\sst(2)}}
\def\DeltaJ{\Delta_s^{\sst({\cal J})}}
\def\calJ{{\cal J}}
\begin{equation}
\Delta_s\ =\ -\partial^2+\Deltaone+\Deltatwo
+\DeltaJ
\label{Deltasdef}
\end{equation}
where in the present model
\begin{equation}
\Deltaone=i\{\partial^\mu\,,\,v_{\bkgd\,\mu}^{\sixd\,a}t^a\}\ ,
\qquad
\Deltatwo=v_{\bkgd\,\mu}^{\sixd\,a}t^a\, v_{\bkgd}^{\sixd\,\mu b}t^b
\ ,\quad
\DeltaJ=v_{\bkgd\,\mu\nu}^{\sixd\,a}t^a\calJ_s^{\mu\nu}
\label{Deltasdef2}
\end{equation}
Here $t^a$ is the color generator in the adjoint representation,
$(t^a)_{bc}=-i\epsilon_{abc},$ 
and $\calJ_s^{\mu\nu}$ is the generator of 6-dimensional Lorentz
transformations in the spin-$s$ representation:
\begin{equation}
\big(\calJ_1^{\mu\nu}\big)_{\lambda\sigma}=i(
\delta^\mu_\lambda\delta^\nu_\sigma-\delta^\mu_\sigma\delta^\nu_\lambda)\ ,
\quad\calJ_{1/2}^{\mu\nu}=\frac{i}{4}[\gamma^\mu,\gamma^\nu]\ ,\quad
\calJ_0^{\mu\nu}=0\ . 
\label{spinsrep}
\end{equation}
We now depart from \cite{peskin} in two obvious ways, in
order to incorporate spontaneous symmetry breaking. First, we require the
background fields to live purely in the third direction in
color space,
\begin{equation}
\uv_{\bkgd\,\mu}^\sixd 
\ =\ v_{\bkgd\,\mu}^{\sixd\,3}\,\tau^3/2\ 
\label{colorspace}
\end{equation}
Second, rather than truncating the expansion of the logarithm at second
order in the total background field, we instead expand about the VEV,
rewriting Eq. (\ref{vevpoint}) as
\begin{equation}
\uv_{\bkgd\,\mu}\ =\ \sqrt{2}\vhiggs\,\delta_{\mu5}\,\tau^3/2
+\delta\uv_{\bkgd\,\mu}
\label{deviation}
\end{equation}
and keep terms to second order in the deviation field
$\delta\uv_{\bkgd\,\mu}$ but to all orders in the VEV itself.
In terms of the deviation field,  Eq. (\ref{Deltasdef}) becomes
\begin{eqnarray}
\Deltaone & = & i\{\partial^\mu\,,\,
\delta v_{\bkgd\,\mu}^{\sixd\,3}t^3\}\ ,
\qquad \nonumber \\
\Deltatwo & = & \Deltatwotilde-2M_W\delta v_{\bkgd\,5}^{\sixd\,3}t^3t^3
-M_W^2t^3t^3
\ ,\quad \nonumber \\
\DeltaJ &= & \delta v_{\bkgd\,\mu\nu}^{\sixd\,3}t^3\calJ_s^{\mu\nu}
\label{Deltasdef3}
\end{eqnarray} 
Here
\begin{equation}
\Deltatwotilde\ =\ \delta v_{\bkgd\,\mu}^{\sixd\,3}t^3\, 
\delta v_{\bkgd}^{\sixd\,\mu 3}t^3
\label{Deltanew}
\end{equation}
Also, thanks to (\ref{colorspace}), 
$\delta v_{\bkgd\,\mu\nu}^{\sixd\,3}$ is now the abelian field strength
$\partial_{[\,\mu}\delta v_{\bkgd\,\nu\,]}^{\sixd\,3}$ subject as always
to the constancy conditions (\ref{constancy}).
Taylor expanding the logarithm then gives
\def\calG{{\cal G}}
\begin{eqnarray}
\sum_{s=0,\frac{1}{2},1}\eta_s\,{\rm Tr}\log\Big(-(\partial^2
+M_W^2t^3t^3)+\Deltaone+\Deltatwotilde+\DeltaJ
-2M_W\delta v_{\bkgd\,5}^{\sixd\,3}t^3t^3\,\Big) & & \nonumber \\
= \ \hbox{const.}+
\sum_{s=0,\frac{1}{2},1}\eta_s\,{\rm Tr}\Big(\,
\calG\cdot\big(\Deltaone+\Deltatwotilde+\DeltaJ
-2M_W\delta v_{\bkgd\,5}^{\sixd\,3}t^3t^3\,\big)  & & \nonumber \\
- 
\calG\cdot\big(\Deltaone+\DeltaJ
-2M_W\delta v_{\bkgd\,5}^{\sixd\,3}t^3t^3\,\big)
\cdot\calG\cdot\big(\Deltaone+\DeltaJ
-2M_W\delta v_{\bkgd\,5}^{\sixd\,3}t^3t^3\,\big)\ \Big)  & &  \nonumber \\
\label{cra}
\end{eqnarray}
where corrections to the RHS are ${\cal O}(\delta v_{\bkgd}^{3})$
Here $\calG$ is the diagonal colour-space matrix, 
\begin{equation}
{\cal G}\ =\ \hbox{diag}\big(\ \frac{1}{ p^2-M_W^2}\,,\,
\frac{1}{ p^2-M_W^2}
\,,\, \frac{1}{p^2}\ \big)
\label{calGdef}
\end{equation}
as follows from $(t^3t^3)_{ab}=\delta_{ab}-\delta_{a3}\delta_{b3}$.
Apart from the masses in the propagators (\ref{calGdef}), the chief effect
of the spontaneous symmetry breaking in (\ref{cra}) are the terms
linear and quadratic in $M_W$. Let us dispose of these first.
Terms linear in $M_W$ are
\begin{eqnarray}
\sum_{s=0,{1\over2},1}\eta_sM_W\,{\rm Tr}\Big(\,
-2\calG\cdot\delta v_{\bkgd\,5}^{\sixd\,3}t^3t^3
+2\calG\cdot\delta v_{\bkgd\,5}^{\sixd\,3}t^3t^3
\cdot\calG\cdot\Deltaone
  && \nonumber \\ \qquad\qquad\qquad\qquad\qquad\qquad
+\ 2\calG\cdot\delta v_{\bkgd\,5}^{\sixd\,3}t^3t^3
\cdot\calG\cdot\DeltaJ\,\Big) &&
\label{linearterms}
\end{eqnarray}
These are would-be tadpole contributions to the effective action.
Respectively, the second and third terms here vanish because
${\rm tr}_{\sst\rm color}t^3t^3t^3=0$ and 
${\rm tr}_{\sst\rm rep}\calJ_s^{\mu\nu}=0$. 
The first term is a nonvanishing Feynman diagram whose only spin dependence
comes from the trace over the Lorentz representation, giving a relative
weight of $6,4,1$, respectively, for the vector, spinor and ghost loop. From 
the above values of $\eta_s$ one sees that
\begin{equation}
6\eta_1+4\eta_{1/2}+\eta_0\ =\ 0
\label{etasum}
\end{equation}
so that the tadpoles do cancel among the three types of
loops (a manifestation of supersymmetry).
And the same arithmetic kills the cross-term in (\ref{cra}) proportional
to $M_W^2.$ (Since these mass-dependent terms were the only potential source
of difference between $v_{\bkgd\,5}^{\sixd\,3}\equiv\phi_{\bkgd 3}^3$
and the other two massless scalars, their vanishing implies that $C_1=C_3$
in the notation of Eq. (\ref{Leffdef}).)
In fact these three arguments kill almost all of the 
$M_W^0$ terms in (\ref{cra}) as well, leaving only
\begin{eqnarray}
\sum_{s=0,\frac{1}{2},1}\eta_s\,{\rm Tr}\,\big(
-\hf\calG\cdot\DeltaJ\cdot\calG\cdot\DeltaJ\big)  = 
-\hf\int{d^3k\over(2\pi)^3}
v_{\mu\nu}^3(k)v_{\rho\sigma}^3(-k) \qquad \qquad \qquad && \nonumber \\ 
 \times\ 
\Big(\int{d^3p\over(2\pi)^3}\,2\cdot{1\over p^2-M_W^2}\cdot
\frac{1}{(p+k)^2-M_W^2}\Big)
\sum_{s=0,\frac{1}{2},1}\eta_s\,{\rm tr}\,{\cal
J}_{s}^{\mu\nu}{\cal J}_{s}^{\rho\sigma} \qquad \qquad && \nonumber \\ 
\label{survive}
\end{eqnarray}
The $p$ integration yields $i/(4\pi M_W)+{\cal O}(k^2)$ 
(the factor of 2 inside
the integrand counts the two massive colors $a=1,2$).
The spin sum simplifies using \cite{peskin} 
\begin{equation}
{\rm tr}\,\calJ_s^{\mu\nu}\calJ_s^{\rho\sigma}\ =\ 
(g^{\mu\rho}g^{\nu\sigma}-g^{\mu\sigma}g^{\nu\rho})C(s)
\label{spinsum}
\end{equation}
where an explicit calculation using (\ref{spinsrep}) gives 
$C(1)=2,$ $C(\hf)=1,$
and $C(0)=0.$ So the net contribution is
\begin{equation}
\frac{i}{8\pi M_W}
\int \,\frac{d^3k}{(2\pi)^3}v_{\bkgd\,\mu\nu}^3(k)
v_{\bkgd}^{3\,\mu\nu}(-k)
\label{netcont}
\end{equation}
where we drop terms of ${\cal O}(k^4)$. A comparison
with the tree-level action 
$-{2\pi \over e^2}{i\over4}\int(v^3_{\mu\nu})^2$ then implies
\begin{equation}
\frac{2\pi}{ e^2}\ \rightarrow\  
\frac{2\pi}{ e^2}-\frac{1}{2\pi M_W}+{\cal O}(e^2)
\label{erenorm}
\end{equation}
which is our one-loop prediction $C_1=C_3=1/4\pi^2$.

\section*{Appendix C: Three-dimensional Instantons}

In this Appendix we set up and give the details of the 1-instanton calculus
in three spacetime dimensions.
\paragraph{}
First we consider the bosonic part of the three-dimensional Euclidean 
action $(\ref{s3d})$,
\begin{equation}
S_{\rm B} ={2\pi \over e^2} \int d^3 x  
{\rm Tr}\big\{\,\hf\uv_{\mu\nu}\uv^{\mu\nu}+
\D_\mu\uphi_i\D^\mu\uphi_i
+\big([\uphi_1,\uphi_2]^2+[\uphi_2,\uphi_3]^2+[\uphi_3,\uphi_1]^2\big)
\big\}
\ .
\label{sb}
\end{equation}
To find the instanton configuration we employ Bogomol'nyi's
lower bound approach \cite{bog},
\begin{eqnarray}
&&S_{\rm B} \ \ge \ 
{2\pi \over e^2}\int d^3x \hf\biggl( B_\mu^a B_\mu^a  + 
 {\cal D}_\mu \phi^a_3 {\cal D}_\mu \phi^a_3 \biggr) \\
&& = {2\pi \over e^2}\int d^3x \ \hf \biggl(
\bigl(B_\mu^a + {\cal D}_\mu \phi^a_3 \bigr)^2  +  
\bigl(-2 B_\mu^a  {\cal D}_\mu \phi^a_3 \bigr)  \biggr) 
 \ge 
{2\pi \over e^2}\int d^3x \bigl(- B_\mu^a  {\cal D}_\mu \phi^a_3 \bigr)
\ , \nonumber
\label{bl}
\end{eqnarray}
where 
$B_\mu^a =\hf\epsilon_{\mu\nu\rho}v^a_{\nu\rho}$.
The 3D bosonic instanton, being a minimum of $ S_{\rm B}$, 
saturates  
Bogomol'nyi bound $(\ref{bl})$,
\begin{equation}
{\uphi^{\rm cl}}_1  =  0 \ , \qquad 
   {\uphi^{\rm cl}}_2  =  0 \ , 
\label{triv}
\end{equation}
\begin{equation}
B_\mu^{{\rm cl} \ a} = \- {{\cal D}^{\rm cl}}_\mu \phi_3^{{\rm cl} \ a}   
\ , 
\label{be}
\end{equation}
where $(\ref{triv})$ ensures the vanishing of the commutator terms in 
$(\ref{sb})$
and the remaining components of the bosonic instanton satisfy
Bogomol'nyi equation $(\ref{be})$ and are given by \cite{PS},
\begin{eqnarray}
\phi_3^{{\rm cl} \ a}  =&& 
\Bigl( M_W |x| \ {\rm coth} M_W |x| -1
\Bigr) \ {x^a \over x^2} 
 \ , \nonumber\\
v_\mu^{{\rm cl} \ a}  =&&  \Bigl( 1-{M_W |x| \over {\rm sinh} M_W |x|}
\Bigr) \ \epsilon_{a\mu\nu}{x^\nu \over x^2} 
 \ ,
\label{bops}
\end{eqnarray}
with the boundary conditions as $|x|\to \infty$,
\begin{equation}
\phi_3^{{\rm cl} \ a} \ \to \ {x^a \over x} M_W \ , \qquad
B_\mu^{{\rm cl} \ a} \ \to \ -{x^a x^\mu \over x^4}
\ .
\label{bc}
\end{equation}
The instanton action follows from $(\ref{bl})$, $(\ref{bc})$,
\begin{equation}
S_{\rm cl}  = \ {2\pi \over e^2}\int d^3x 
\partial_\mu\bigl(-B_\mu^{{\rm cl} \ a} \phi_3^{{\rm cl} \ a}\bigr) = \
{2\pi \over e^2} \ 4\pi M_W
\label{bact}
\end{equation}
From the above formulae it is obvious that the bosonic
components of the 3D instanton are those of the BPS (anti-)monopole
in the corresponding four-dimensional theory in the $\uv_0=0$ gauge.
\paragraph{}
The fermi-field components of the instanton can be determined by
infinitesimal supersymmetry transformations of the bosonic
components $(\ref{bops})$ and will be discussed later.
\paragraph{}
Isolated one-instanton contribution 
to the functional integral is of the generic
form \cite{th},
\begin{equation}
{\cal Z}_1 \ = \ \int \ d\mu_{B} \ \int \ d\mu_{F} \ R \ \exp[-S_{\rm cl}]
\ ,
\label{z1}
\end{equation}
where $d\mu_{B}$ and $d\mu_{F}$ are the measures of integrations over
collective coordinates of bosonic and fermionic zero modes, $R$ is
the ratio of functional determinants over non-zero eigenvalues
of the operators of quadratic fluctuations in the instanton background,
Eq. $(\ref{cancel})$. In this Appendix we  determine $d\mu_{B}$
and $d\mu_{F}$.
\paragraph{}
It will prove to be particularly convenient to arrange the 3D instanton
analysis in a four-dimensional way. The four-vector field
$\uv_m$ of the dimensionally reduced 4D theory 
(translationally invariant along $x_3$) is given by $(\ref{df3})$.
The Bogomol'nyi equations $(\ref{be})$ in this language are 
equivalent to the 4D self-duality equations \cite{lohe},
\begin{equation}
v_{m n}^{{\rm cl} \ a} \ {= \ }^* v_{m n}^{{\rm cl} \ a}
\label{selfd}
\end{equation}
and the instanton action $(\ref{bact})$ is
\begin{equation}
S_{\rm cl} \ = \ {2\pi \over e^2}\int d^3x \ 
\quarter v_{m n}^{{\rm cl} \ a} v_{m n}^{{\rm cl} \ a} \ = \
{8\pi^2 \over e^2}  M_W
\ .
\label{sacl}
\end{equation}
The 3D instanton or, equivalently, the BPS monopole, is in this way
similar to the 4D Yang-Mills instanton \cite{BPST}.
The functional integration over the 
($x_3$-independent) fluctuations $\delta v_{m}^{a}$
around the instanton will be performed
in the (four-dimensional) covariant background gauge,
\begin{equation}
{\cal D}_m^{\rm cl} \delta v_{m}^{a} \ = \ 0
\ ,
\label{bkgg}
\end{equation}
and the bosonic instanton zero modes will be of the general
form \cite{Bernard},
\begin{equation}
 Z_m^{a \ [k]} \ = \ {\partial v_{m}^{{\rm cl} \ a}
\over \partial \gamma^{[k]}} \ + \ {\cal D}_m^{\rm cl} \Lambda^a
\ ,
\label{zm1}
\end{equation}
where $\gamma^{[k]} $ are the zero modes collective coordinates --
the three-translations, $X_\mu$, and the $U(1)$ rotation, $\theta$.
The second term on the right hand side of $(\ref{zm1})$ is necessary
to keep $ Z_m$ in the background gauge $(\ref{bkgg})$.
In general, these bosonic zero modes, $Z_{m}$, can be written 
in a more compact notation,
\begin{equation}
{\cal D}^{\rm cl}_{[m}Z_{n]}=\,^{*}{\cal D}^{\rm cl}_{[m}Z_{n]}
\ \ \ ,\ \ \ {\cal D}^{{\rm cl} \ m}Z_{m}=0 \ .
\label{fancy}
\end{equation}
The measure $d\mu_{B}$ is now simply \cite{Bernard},
\begin{equation}
d\mu_{B} \ = \ \prod_{k=1}^{4} {d \gamma^{[k]} \over \sqrt{2\pi}} \ 
\biggl( {\rm det}_{rs} \ {2\pi \over e^2}\int d^3x \  
Z_m^{a \ [r]} \ Z_m^{a \ [s]} \biggr)^{1/2}
\ .
\label{mesb}
\end{equation}
First, consider translational zero modes, cf. $(\ref{zm1})$,
\begin{equation}
Z_m^{a \ [\nu]} \ = \ -\partial_\nu v_{m}^{{\rm cl} \ a}
\ + \ {\cal D}_m^{\rm cl} v_{\nu}^{{\rm cl} \ a} \ = 
v_{m \nu}^{{\rm cl} \ a}
\ .
\label{tra}
\end{equation}
Their overlap is
\begin{eqnarray}
{\cal O}_{\mu\nu}  &=& {2\pi \over e^2}\int d^3x \ Z_m^{a \ [\mu]} \ 
Z_m^{a \ [\nu]} \ = \ {2\pi \over e^2}\int d^3x \ 
v_{m \mu}^{{\rm cl} \ a} \ v_{m \nu}^{{\rm cl} \ a} \nonumber\\
 &&= {\delta_{\mu \nu} \over 4} {2\pi \over e^2}\int d^3x \ 
v_{mn}^{{\rm cl} \ a} \ v_{mn}^{{\rm cl} \ a} \ = \ 
\delta_{\mu \nu} \  S_{\rm cl}
\ .
\label{trov}
\end{eqnarray}
Now we consider the $U(1)$ orientation zero mode.
To do this we need to, first,  gauge rotate the instanton $(\ref{bops})$
into the unitary (singular) gauge, where 
\begin{equation}
v_{0 \ {\rm sing}}^{\rm cl \ a}\ = \ 
\phi_{3 \ \rm sing}^{\rm cl \ a} \sim \delta^{a3}
\ ,
\label{sin}
\end{equation}
and, second, allow the
further global gauge transformations consistent with $(\ref{sin})$,
they obviously form a $U(1)$ subgroup of the $SU(2)$,
\begin{equation}
{\tilde V}_{m \ {\rm sing}}^{\rm cl} \ = \
\exp[i \theta \tau^3/2] \ {\tilde v}_{m \ {\rm sing}}^{\rm cl} 
\ \exp[- i \theta \tau^3/2] 
\ .
\label{u1s}
\end{equation}
The $U(1)$ zero mode in the form  $(\ref{zm1})$ is 
\begin{equation}
{\tilde Z}_m^{[3]} \ = \ 
\partial_{\theta} {\tilde V}_{m \ {\rm sing}}^{\rm cl} 
\ + \ {1 \over M_W} \ {\cal D}_m^{\rm cl} \ \Bigl(
{\tilde \phi}_{\rm sing}^{\rm cl} - M_W \tau^3 / 2 \Bigr)
\ = \ {1 \over M_W} \ {\tilde v}_{m 3 \ {\rm sing}}^{\rm cl} \ ,
\label{gu1}
\end{equation}
with the overlaps,
\begin{eqnarray}
{\cal O}_{33}  &=& {2\pi \over e^2}\int d^3x \ Z_m^{a \ [3]} \ 
Z_m^{a \ [3]} \ = \ {1 \over M_W} \ {2\pi \over e^2}\int d^3x \ 
v_{m 3}^{{\rm cl} \ a} \ v_{m 3}^{{\rm cl} \ a} \ = \ 
{1 \over M_W} \ S_{\rm cl} \nonumber\\
&& {\cal O}_{\mu 3} =  {2\pi \over e^2}\int d^3x \ Z_m^{a \ [\mu]} \ 
Z_m^{a \ [3]} \ = \ 0
\ .
\label{fover}
\end{eqnarray}
Finally combining $(\ref{mesb})$, $(\ref{trov})$ and $(\ref{fover})$
we obtain the desired expression $(\ref{bmeasure})$ for $d\mu_B$, 
\begin{equation}
\int d\mu_B \ = \ \int {d^3 X \over (2 \pi)^{3/2}} \ S_{\rm cl}^{3/2}
\int_{0}^{2\pi} {d\theta \over (2 \pi)^{1/2}}  \ 
{S_{\rm cl}^{1/2}\over M_W}
\ .
\label{dmbtot}
\end{equation}
\paragraph{}
Our next goal is $d\mu_{F}$. The 1-instanton solution of the
$N=4$ supersymmetric Yang-Mills in 3D has four fermion zero modes
which can be determined by
infinitesimal $N=4$ supersymmetry transformations of the bosonic
components $(\ref{bops})$. Equivalently they can be understood 
in terms of the Weyl spinors of the four-dimensional theory, 
$(\ref{fmodes})$,
\begin{eqnarray}
\lambda^{\rm cl}_{\alpha} & =& \hf\xi_{\beta}
(\sigma^{m}\sigmabar^{n})_{\alpha}^{\ \beta}
v^{\rm cl}_{mn} \nonumber \\ 
\psi^{\rm cl}_{\alpha} & =& \hf \xi'_{\beta}
(\sigma^{m}\sigmabar^{n})_{\alpha}^{\ \beta}
v^{\rm cl}_{mn} \ ,
\label{fmodes2} 
\end{eqnarray}
where the two-component
Grassmann collective coordinates $\xi_{\alpha}$ and $\xi'_{\alpha}$
are the parameters of infinitesimal $N=2$ supersymmetry 
transformations in (dimensionally reduced) 4D theory.
Since $\alpha=1,2$ there are two $\lambda$'s and two $\psi$'s 
which is equivalent to four zero modes in terms of
Majorana spinors in 3D.
There are no anti-fermion zero modes due to the self-duality,
$(\ref{selfd})$.
The fermion collective coordinates integration measure is
in general, 
\begin{equation}
\int \, d\mu_{F}=\int\, d^{2}\xi\, d^{2}\xi' \ ({\cal J}_{\xi})^{-2} \ ,
\label{f2measure}
\end{equation}
where the fermion Jacobian is,
\begin{equation}
{\cal J}_{\xi}={2\pi \over e^2} \int d^3 x \ d^{2}\xi\, 
{\rm Tr} \ \ulambda^{{\rm cl} \ \alpha} \ulambda^{\rm cl}_{\alpha} 
\ ,
\label{fjack}
\end{equation}
and it is understood\footnote{Note that in $(\ref{fjack})$ 
we could have summed over the isospin
$a=1,2,3$ rather than trace over the $SU(2)$ matrices.
This would have produced an extra factor of $1/2$ which then
would require an alternative prescription for Grassmanian
integration, $\int d^{2}\xi\, \xi^2 \equiv 2$ and the final answer
for ${\cal J}_{\xi}$
would be the same.}  
that $\int d^{2}\xi\, \xi^2 \equiv 1$. 
The right hand side of $(\ref{fjack})$ is easily evaluated
with the use of $(\ref{fmodes2})$ and the sigma-matrix algebra,
and gives ${\cal J}_{\xi} = 2S_{\rm cl}$.
Comparing with $(\ref{fmeasure})$ we obtain, 
\begin{equation}
\int \, d\mu_{F}=\int\, d^{2}\xi\, d^{2}\xi' \ (2S_{\rm cl})^{-2} \ .
\label{f22measure}
\end{equation}
\paragraph{}
Finally we need the long distance asymptotics of the
fermion zero modes $(\ref{fmodes2})$. 
These can be readily obtained by, first, switching 
from $\sigma^m$ matrices in $(\ref{fmodes2})$ to $\Gamma_{\rm ch}^m$
of Eq. $(\ref{gch})$, second, rotating into the real basis,
$(\ref{gre})$, decoupling the $x_2$ direction and, finally,
switching to $\gamma^\mu$ matrices in 3D.
The result is then rotated into the singular gauge, 
$(\ref{sin})$, and the large distance limit
$|x|\to \infty$ is considered
\begin{equation}
\lambda^{\rm LD}_{\alpha} (x)=  2\gamma_{\alpha}^{\mu \ \beta} \ \xi_{\beta} 
\ {x_\mu \over x^3} \ , \quad
\psi^{\rm LD}_{\alpha} (x) =  \ 2\gamma_{\alpha}^{\mu \ \beta}\ {\xi'}_{\beta} 
\ {x_\mu \over x^3} \ ,
\label{llld}
\end{equation} 
which confirms Eq. $(\ref{ld})$.

\section*{Appendix D: The Riemann Tensor}

In this appendix we calculate the leading exponentially suppressed terms in  
the Riemann tensor for the metric

\[
ds^{2}=f^{2}(r)dr^{2}+a^{2}(r)\sigma_{1}^{2}+b^{2}(r)\sigma_{
2}^{2}+c^{2}
(r)\sigma_{3}^{2} \]
with $\sigma_{i}$ defined in (\ref{1forms}) and the functions $a, b$ and $c$ 
satisfying (\ref{ode}). Following Appendix B of \cite{gh}, 
we define the vierbein one-forms 
$\hat{\theta}^{\alpha}=\theta^{\alpha}_{i}dX_{i}$ with
\begin{equation}
\hat{\theta}^{1}=a\sigma_{1}\ 
\ ,\ \ \hat{\theta}^{2}=b\sigma_{2}\ \ ,\ \ \hat{\theta}^{3}=c\sigma_{3} 
\ \,\ \  \hat{\theta}^{4}=fdr 
\label{vier}
\end{equation}
which, using (\ref{ode}) gives us the spin connection 
\begin{equation}
 \omega^{1}_{r}=(a'/f)\sigma_{1}\ ,\ 
\omega^{2}_{r}=(b'/f)
\sigma_{2}\ ,\ \omega^{3}_{r}=(c'/f)\sigma_{3} \]
\[ \omega^{1}_{2}=(1+c'/f)\sigma_{3}\ ,\ 
\omega^{2}_{3}=(1+a'/f)\sigma_{1}\ ,\ 
\omega^{3}_{1}=(1+b'/f)\sigma_{2}
\label{spinconn}
\end{equation}
The curvature 2-forms are now found to be 
\begin{eqnarray} 
R^{12}= \hat{c}'dr\wedge\sigma_{3} + [-\hat{c}+\hat{a}+\hat{b}+2\hat{a}
\hat{b}]\sigma_{1}\wedge\sigma_{2} \nonumber \\
R^{23}= \hat{a}'dr\wedge\sigma_{1} + [-\hat{a}+\hat{b}+\hat{c}+2\hat{b}
\hat{c}]\sigma_{2}\wedge\sigma_{3} \nonumber \\
R^{31}= \hat{b}'dr\wedge\sigma_{2} + [-\hat{b}+\hat{c}+\hat{a}+2\hat{c}
\hat{a}]\sigma_{3}\wedge\sigma_{1} 
\end{eqnarray}
Where $\hat{a}=a'/f$, $\hat{b}=b'/f$ and $\hat{c}=c'/f$. The 
other components are determined by $R^{34}=R^{12}$ 
and cyclic, giving us a Riemann tensor self-dual with respect to 
the epsilon tensor $\epsilon^{1234}=+1$. Explicitly we have,
\begin{equation}
R_{\alpha\beta\gamma\delta}=\eta^{a}_{\alpha\beta}\,T_{ab}\,\eta
^{b}_{\gamma\delta}
\label{rt1}
\end{equation}
with $T={\rm diag}(-\hat{a}'/fa,-\hat{b}'/fb,-\hat{c}'/fc)$ where,  
\begin{equation}
R_{ijkl}=\theta^{\alpha}_{i}\theta^{\beta}_{j}\theta^{\gamma}_{k}
\theta^{\delta}_{l} R_{\alpha\beta\gamma\delta}
\label{rt0}
\end{equation}
Calculating the Riemann tensor at the point 
$(0,0,r,\sigma)$ in the standard coordinates,  we then change to the 
complex basis (\ref{cbasis}) with coordinates
\begin{eqnarray}
z_{1}=\frac{1}{\sqrt{2}}(X-iY) & \qquad{} \qquad{} \qquad{} & 
z_{2}=\frac{1}{\sqrt{2}}(Z- i\sigma)
\end{eqnarray}
Labelling the coordinates $(z_{1},z_{2},\bar{z}_{1},\bar{z}_{2})$ by an 
index $P=1,2,3,4$, the Riemann tensor is conveniently given in terms of the 
following $(4\times 4)$ matrices,  
\begin{eqnarray}
 A & = & {\small\left(\begin{array}
{cccc} 0 & A_{+}e^{+i\psi} & 0 & A_{-}e^{+i\psi} \\ -A_{+}e^{+i\psi} 
& 0 & -A_{-}e^{-i\psi} & 0 \\ 0 & A_{-}e^{-i\psi} & 0 & A_{+}e^{-i\psi} \\ 
-A_{-}e^{+i\psi} & 0 & -A_{+}e^{-i\psi} & 0 \end{array}
\right)} \nonumber \\ B & = &  {\small\left
(\begin{array}{cccc}0 & -B_{+}e^{+i\psi} & 0 & -B_{-}e^{+i\psi} 
\\ B_{+}e^{+i\psi} & 0 & -B_{-}e^{-i\psi} & 0 \\ 0 & B_{-}e^{-i\psi} & 0 
& B_{+}e^{-i\psi} \\ B_{-}e^{+i\psi} & 0 & -B_{+}e^{-i\psi} & 0 
\end{array}\right)} \nonumber \\ C  & = & 
{\small\left(\begin{array}{cccc} 0 & 0 & \frac{ab}{r^{2}}
 & 0 \\ 0 & 0 & 0 & \frac{cf}{2} \\ -\frac{ab}{r^{2}} & 0 & 0 & 0 \\ 0 & 
-\frac{cf}{2} & 0 & 0 \end{array}\right)}
\label{rt2}
\end{eqnarray}
where $A_{\pm}=af\pm\frac{1}{2}bc$ and $B_{\pm}=bf\pm\frac{1}{2}ac$. The final result is  
\begin{eqnarray}
R_{PQRS} & = & -\frac{\hat{a}'}{4far^{2}} A_{PQ}A_{RS} +
\frac{\hat{b}'}{4fbr^{2}}B_{PQ}B_{RS} +\frac{\hat{c}'}{fc}C_{PQ}C_{RS}
\label{rt4}
\end{eqnarray}
\paragraph{}
The leading exponentially suppressed terms are proportional to $(a-b)$ whose 
asymptotic form is given in (\ref{prediction}) as, 
\begin{eqnarray}
a-b & \simeq & -8qr^{2}e^{-r}
\end{eqnarray} 
Hence, to compare with the instanton calculation, 
it suffices to approximate all other factors by their 
leading weak coupling behaviour. From (\ref{abc}) we have,  
\begin{eqnarray}
a+b & \simeq & 2r+O(1/r) \nonumber   \\
c & \simeq & -2+O(1/r) 
\nonumber \\ f & \simeq & -1+O(1/r) 
\label{asymptotic3}
\end{eqnarray}
Expanding the exact expression (\ref{rt4}) we find that, to the required 
order, 
\begin{eqnarray}
 R_{1212} &= & 8qre^{-r+i\sigma} \nonumber \\ 
 R_{\bar{1}\bar{2}\bar{1}\bar{2}} & = & 8qre^{-r-i\sigma} 
\label{rt3}
\end{eqnarray}
As discussed in Section 4, the remaining  pure (anti)-holomorphic 
components are related to these by symmetries of the Riemann tensor. All 
components of mixed holomorphy are independent of $\sigma$.

\end{document}